\documentclass[a4paper,aps,pra,twocolumn,superscriptaddress,showpacs]{revtex4}

\usepackage[utf8]{inputenc}
\usepackage{amssymb}
\usepackage{amsmath}
\usepackage{multirow}

\usepackage[pdftex]{graphicx}
\usepackage[pdftex]{color}
\usepackage[T1]{fontenc}

\renewcommand{\i}{{\rm i}}
\renewcommand{\d}{{\rm d}}

\renewcommand{\O}[2][]{\hat{#2} ^{\vphantom{\dagger}}_{#1}}
\newcommand{\Op}[2][]{\hat{#2} ^{\dagger}_{#1}}

\begin{document}

\title{Bosons in a 2d bichromatic quasi-periodic potential: Analysis of the disorder in the Bose-Hubbard parameters and phase diagrams}

\author{A.\,E.\,Niederle and H.\,Rieger}

\address{Theoretical Physics, Saarland University , D--66041 Saarbr\"ucken, Germany}

\begin{abstract}
Experimental realizations of disorder in optical lattices generate a distribution of the Bose-Hubbard (BH) parameters, like on-site potentials, hopping strengths, and interaction energies. We analyze this distribution for bosons in a 2d bichromatic quasi-periodic potential by determining the generalized Wannier functions and calculating the corresponding BH parameters. Using a local mean-field cluster analysis, we study the effect of the corresponding disorder on the phase diagrams. We find a substantial amount of disorder in the hopping strengths, which produces strong deviations from the phase diagram of the disordered BH model with purely random on-site potentials.

\end{abstract}

%\pacs{00.00, 20.00, 42.10}

\maketitle

\section{Introduction}\label{section:Introduction}
Bosons in a regular optical lattice, which can be described by the BH model, show a transition from an incompressible, non-coherent Mott insulator (MI) phase to a compressible, coherent superfluid (SF) phase, which has been shown experimentally \cite{Grei02} as well as theoretically \cite{Fish89, Oost01}. In the disordered case the Bose-glass (BG) phase occurs between the MI and the SF, which is compressible, but not coherent and whose existence has been shown experimentally \cite{Fall07, Meld15}.\\
Theoretically, disordered BH systems are modeled with isolated on-site box disorder, while the other parameters have precise values. Numerical methods like quantum Monte Carlo (QMC) methods \cite{Soey11,Lin11,Capo08,Poll09,Prok04,Lee01,Kisk97,Kisk97a,Krau91} and density-matrix renormalization group techniques \cite{Roux08, Deng08, Deng09,Carr10,Raps99} have been applied to this system in order to study the phase diagram. Based on the mean-field approximation \cite{Shes93} various numerical techniques, such as local mean-field (LMF) theory \cite{Buon09,Buon07,Buon04b}, stochastic mean-field theory \cite{Biss09, Biss10}, and LMF cluster analysis \cite{Nied13} have been proposed. The phase diagram of the 2d disordered BH model with random on-site potentials predicted by the LMF cluster analysis \cite{Nied13}, which interprets the BG-SF transition as a percolation of SF regions, agrees very well with the one obtained from QMC methods \cite{Poll09}. The authors of \cite{Stas14} apply the percolation criterion for their studies on the BH model with random impurities.\\
Disorder in other BH parameters, like the tunneling rates or the inter-particle interactions, has not been studied as intensively as disorder in the on-site potentials. Disorder in the inter-particle interaction can be realized experimentally near a Feshbach resonance \cite{Gimp05,Wild05}. A uniform distribution of this parameter has been studied in \cite{Gimp05,Gimp06}. Disorder of the tunneling rates has been studied within SMF theory \cite{Biss10} for a uniform distribution. Other works on disordered tunneling rates focus on bimodal distributions.\\
Experimentally, disorder can be introduced either by a speckle field \cite{Whit09} or by a bichromatic quasi-periodic lattice produced by two lasers with incommensurate wave lengths \cite{Lye05,Fall07}. While the distributions of the BH parameters resulting from speckle fields have been studied in \cite{Whit09, Zhou10}, in this paper we focus on bosons in two-dimensional bichromatic quasi-periodic potentials and analyze the distribution of all of the BH parameters, i.e. on-site potential, hopping strength and interaction energies, by determining the generalized Wannier functions and calculating the corresponding BH parameters.\\
This paper is organized as follows: First, we discuss the BH model and the LMF cluster analysis in section \ref{section:model}. In section \ref{scenarios} we determine the phase diagrams of the disordered BH model for three different cases of uncorrelated disorder for comparison: exclusively random on-site potentials, exclusively random hopping strengths and exclusively random interaction energies. In section \ref{quasi-periodic} the BH parameters are calculated for a two-dimensional bichromatic quasi-periodic potential and their distributions are characterized. These are finally used in section \ref{Phase diagrams of the bichromatic potential} to determine the phase diagrams. The paper concludes with a discussion.

\section{The Bose-Hubbard model}\label{section:model}
The BH Hamiltonian describing Bosons in an optical lattice \cite{Jaks98} is given by
\begin{equation}\label{BHM}
  \O{H}=-\mu \sum_i  \O[i]{n}+\frac{U}{2} \sum_i \O[i]{n} \left(\O[i]{n}-1\right)-J \sum_{\langle i,j\rangle} \Op[i]{a} \O[j]{a}.
\end{equation}
The operator $\O[i]{n}=\Op[i]{a}\O[i]{a}$ is the particle number operator of bosons on site $i$, which are annihilated and created by the operators $\O[i]{a}$ and $\Op[i]{a}$. The site index $i=1,\ldots,M$, where $M=L^2$ is the number of sites in a $L\times L$ 2d lattice, represents a tuple of spatial coordinates $\left(i_x,i_z\right)$ with $i_{x,z}=1,\ldots,L$. The chemical potential is denoted by $\mu$, the inter particle repulsion by $U$ and the tunneling rate by $J$. The last sum runs over all four ($Z=4$) nearest neighbor pairs $\langle i,j\rangle$ of the lattice.\\
In the ordered case, the $\mu/U$-$J Z/U$ phase diagram displays the well-known Mott-lobes with boundaries given in LMF theory by \cite{Fish89,Oost01}
\begin{eqnarray}\label{Oosten}
 \mu_\pm\left(J,Z,U,n\right)&=&-\frac{1}{2}\left(JZ-U\left(2 n-1\right)\right)\nonumber \\
 &\phantom{=}&\pm\sqrt{\frac{1}{4}\left(JZ-U\right)^2-JZUn}.
\end{eqnarray}
Inside the Mott-lobes the particle number is fixed to an integer value $n$. They are aligned along the $\mu$-axis and the particle number increases from lobe to lobe with growing $\mu$. In this region particle tunneling is prohibited due to the existence of an energy gap in the particle excitation spectrum. The Mott-lobes are surrounded by the SF phase, where particle tunneling is favorable and coherence grows with increasing tunneling rate.\\
In the case of on-site disorder, the BG phase occurs between the MI and SF phase. In this phase the system is compressible, but not SF. In the present paper we will not restrict to on-site disorder, but rather study the influence of disorder on all BH parameters. Therefore, all BH parameters, except the chemical potential $\mu$, which is a global parameter fixing the particle number, now become site dependent: 
\begin{equation}\label{BHMDis}
  \O{H}=\sum_i \left(\epsilon_i-\mu\right) \O[i]{n}+\frac{U_i}{2} \sum_i \O[i]{n} \left(\O[i]{n}-1\right)-\sum_{\langle i,j\rangle} J_{ij} \Op[i]{a} \O[j]{a}.
\end{equation}
The SF parameter   
\begin{equation}\label{SFP}
 \psi_i=\langle\O{a_i} \rangle 
\end{equation}
is defined as the expectation value of the annihilation operator and can be chosen to be real because of the $U\left(1\right)\textendash$symmetry of the BH-Hamiltonian. With the help of the LMF approximation \cite{Shes93}
\begin{equation}
  \O[i]{a} \Op[j]{a}\approx \O[i]{a} \psi_j + \Op[j]{a} \psi_i-\psi_i \psi_j,
\end{equation}
the Hamiltonian can be transformed into a sum of local Hamiltonians $\O{H}=\sum_i \O[i]{H}$,
\begin{equation}\label{HLMF}
  \O[i]{H}=\left(\epsilon_i-\mu\right) \O[i]{n}+\frac{U_i}{2} \O[i]{n} \left(\O[i]{n}-1\right)-\eta_i \left(\O[i]{a}+\Op[i]{a}-\psi_i\right),
\end{equation}
which are effectively coupled by a local hopping rate $\eta_i:=\sum_j J_{ij} A_{ij} \psi_j$, which depends on the local SF parameter of the neighboring sites with $A_{ij}=1$ for $i$ and $j$ being nearest neighbors on the square lattice with periodic boundary conditions and zero otherwise. One should note that mean field approximations as the one used here neglect spatial correlations of quantum fluctuations, which is less severe in high space dimensions. Actually it is exact in infinite dimensions and reproduces the exact critical behavior already above the upper critical dimension, which is unknown for the system we consider here but certainly larger than $4$. Consequently, we expect the approximation to be critically inaccurate for 1d systems and therefore, focus on two space dimensions. We would expect the predictions of our LMF analysis to be even more accurate for three dimensions, but 3d systems are computationally much more demanding.\\
Since the Hamiltonian itself depends on the SF parameter \cite{Buon07}, the self-consistency equation \eqref{SFP} is solved recursively in order to find the ground state of the LMF Hamiltonian \eqref{HLMF}. In turn all local parameters of interest, especially the local particle number $n_i=\langle \Op[i]{a} \O[i]{a}  \rangle$, can be computed. In order to determine the phase boundaries, several steps are needed \cite{Nied13}: First, we define and identify so-called MI and SF sites. Sites with an integer number of particles are called MI sites. Note that the LMF approximation neglects quantum fluctuations and predicts an integer expectation value for the local particle number for some site even in the presence of disorder in the BH parameters. Sites with non-integer expectation value of the local particle number are denoted as SF sites. In a second step all three occurring phase are identified: In the MI phase the system only consists of MI sites and no SF site occurs. The BG is characterized by a mixture of MI and SF sites, more precisely by isolated clusters (islands) of SF sites within a sea of MI sites. In \cite{Nied13} we discussed how this definition is physically plausible regarding the conventional hallmarks of the BG phase, namely the lack of coherence and a gapless spectrum. Since the particle number fluctuates within the isolated SF clusters only the sites within a single cluster can be phase coherent, sites in different clusters are not, which establishes the lack of macroscopic phase coherence. Moreover, the BG phase is actually the Griffiths phase of the BH model (see \cite{Nied13} for a discussion), where the isolated SF clusters behave like finite systems within the SF phase and therefore, have a very small gap that decreases quickly with the size of the SF cluster. Since SF clusters can be arbitrarily large there is no lower bound for the gap and thus the BG phase is gapless. While approaching the BG-SF transition the regions with SF site grow, form connected clusters, which finally percolate. The percolation of the SF sites marks the transition to the SF phase. We call this scheme the LMF cluster analysis approach, which is described and discussed in \cite{Nied13}. In contrast to other LMF approaches we show, that the LMF cluster analysis reproduces the phase diagram predicted by quantum Monte Carlo \cite{Soey11} with excellent accuracy.

\section{Box distributed disorder and different scenarios}\label{scenarios}
Before we focus on the bichromatic quasi-periodic potential, we analyze the effect of uncorrelated disorder in each of the BH parameter $\epsilon_i$, $J_{ij}$ and $U_i$ separately. Here we determine, with the LMF cluster analysis, the phase diagram for a uniform distribution \mbox{$p(\alpha)=\Theta\left(\Delta_\alpha/2-|\alpha|\right)/\Delta_\alpha$} for each BH parameter \mbox{$\alpha=\epsilon_i, J_{ij},U_i$} separately. While all three phases can be found in each disorder scenario, we find substantial differences in the phase diagrams.
\subsection{Disorder in $\epsilon$}\label{section:DissEps}
The most common disorder scenario is diagonal disorder introduced by site-dependent local on-site energies $\epsilon_i$, which are drawn from a box distribution \mbox{$p(\epsilon_i)=\Theta\left(\Delta_{\epsilon}/2-|\epsilon_i|\right)/\Delta_{\epsilon}$}. This has been widely studied via quantum Monte-Carlo methods \cite{Soey11, Capo08,Kisk97, Lin11}, mean-field techniques \cite{Buon07, Biss09, Nied13} and analytic approaches \cite{Free96, Poll09}.\\
\begin{figure}[!htb]
a)\includegraphics[width=3.9cm]{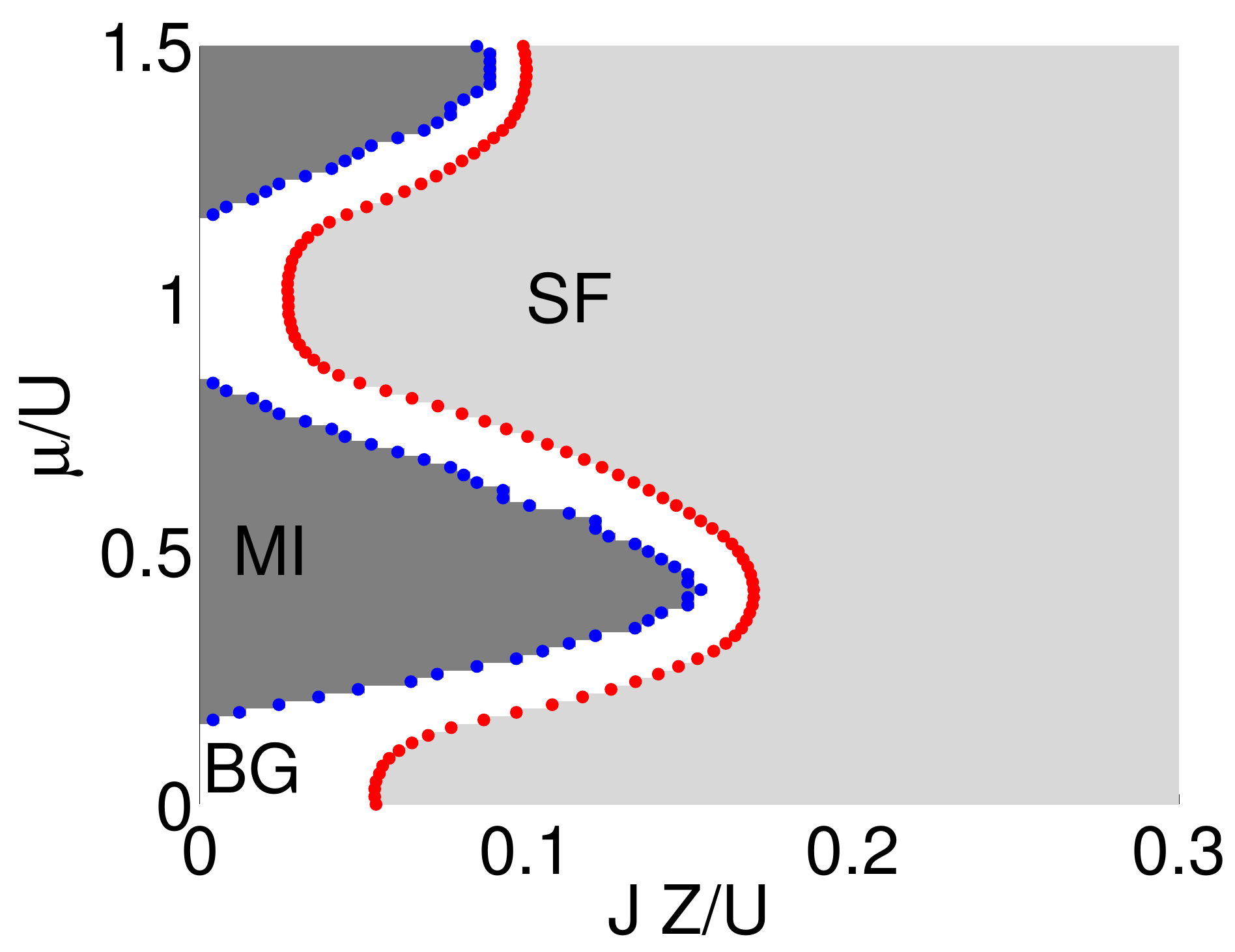}
b)\includegraphics[width=3.9cm]{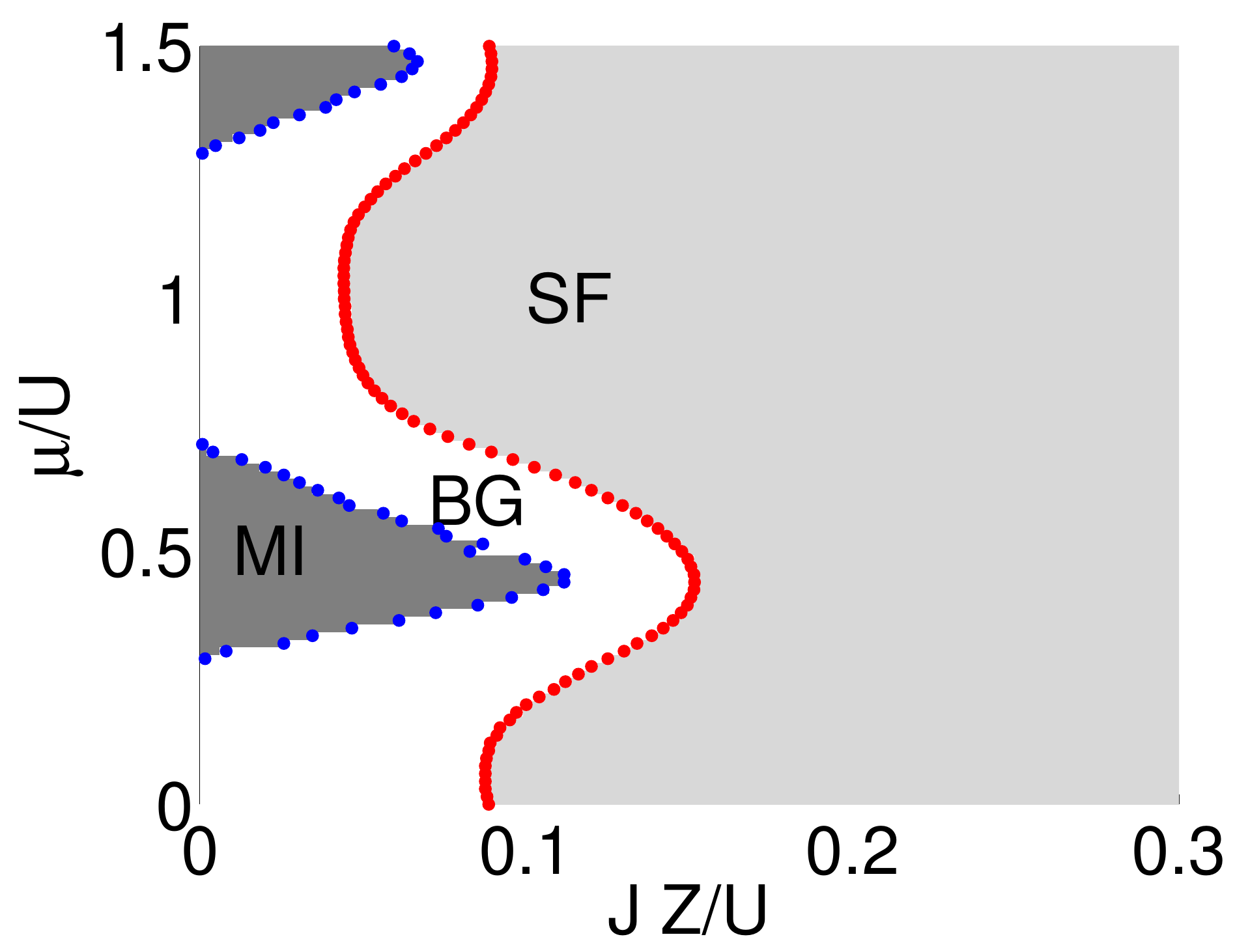}
c)\includegraphics[width=3.9cm]{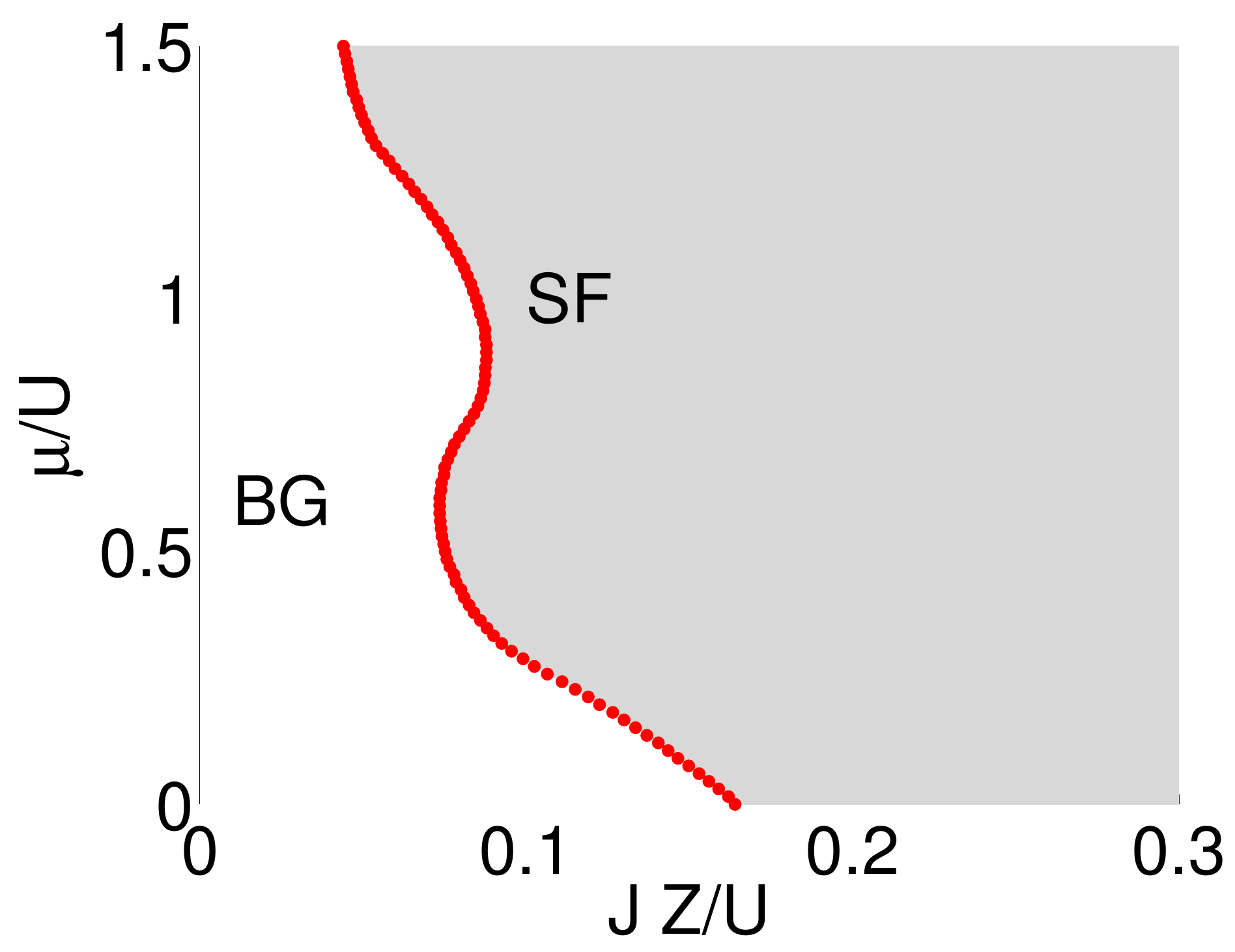}
{\caption{\label{PhasenEpsilon} Phase diagram for box distributed disordered on-site energies $\epsilon_i$ with the disorder strength $\Delta_\epsilon/U=0.35,\,0.6,\,1.5$.}}
\end{figure}
Figure \ref{PhasenEpsilon} shows the phase diagram for different disorder strengths resulting from the LMF-Cluster-Analysis \cite{Nied13}. The Mott-lobes with a fixed particle number of $n=\langle \O[i]{n}\rangle_{av}$ extend from \mbox{$\mu_-=\left( n-1\right) U +\Delta_\epsilon/2$} to \mbox{$\mu_+=n U-\Delta_\epsilon/2$}, see \cite{Fish89}. Thus, all Mott-lobes simultaneously disappear at a critical disorder strength of $\Delta_\epsilon^c/U=1$. The Mott-lobes are surrounded by a BG region. For larger tunneling rates a phase transition to the SF regime occurs. While all three phases appear for $\Delta_\epsilon^c/U<1$, in the strong disorder limit only the BG in the small tunneling and the SF phase in the high tunneling regime survive. 

\subsection{Disorder in $J$}\label{section:DissJ}
The influence of disordered tunneling rates has mainly been studied for bimodal distributions, where two values of the tunneling rate are chosen and distributed randomly among the lattice \cite{Seng07, Prok04, Bala05,Biss10} leading to fundamentally different phase diagrams compared to the one discussed here. In contrast, here we focus on a general approach, where the local tunneling rates are uniformly distributed according to \mbox{$p(J_{ij})=J+\Theta\left(\Delta_J/2-|J_{ij}|\right)/\Delta_J$} symmetric around a fixed value $J$.\\
\begin{figure}[!htb]
a)\includegraphics[width=3.9cm]{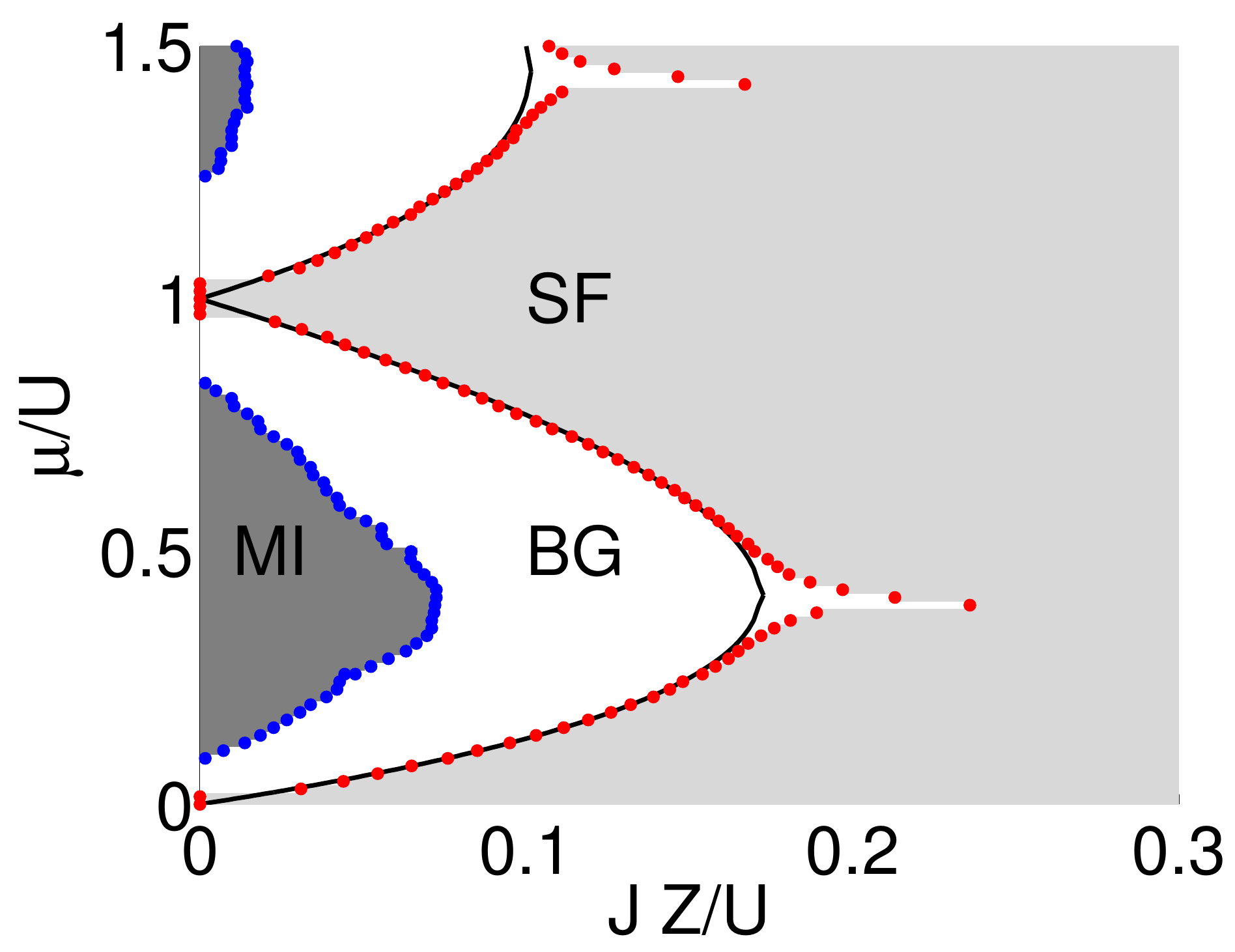}
b)\includegraphics[width=3.9cm]{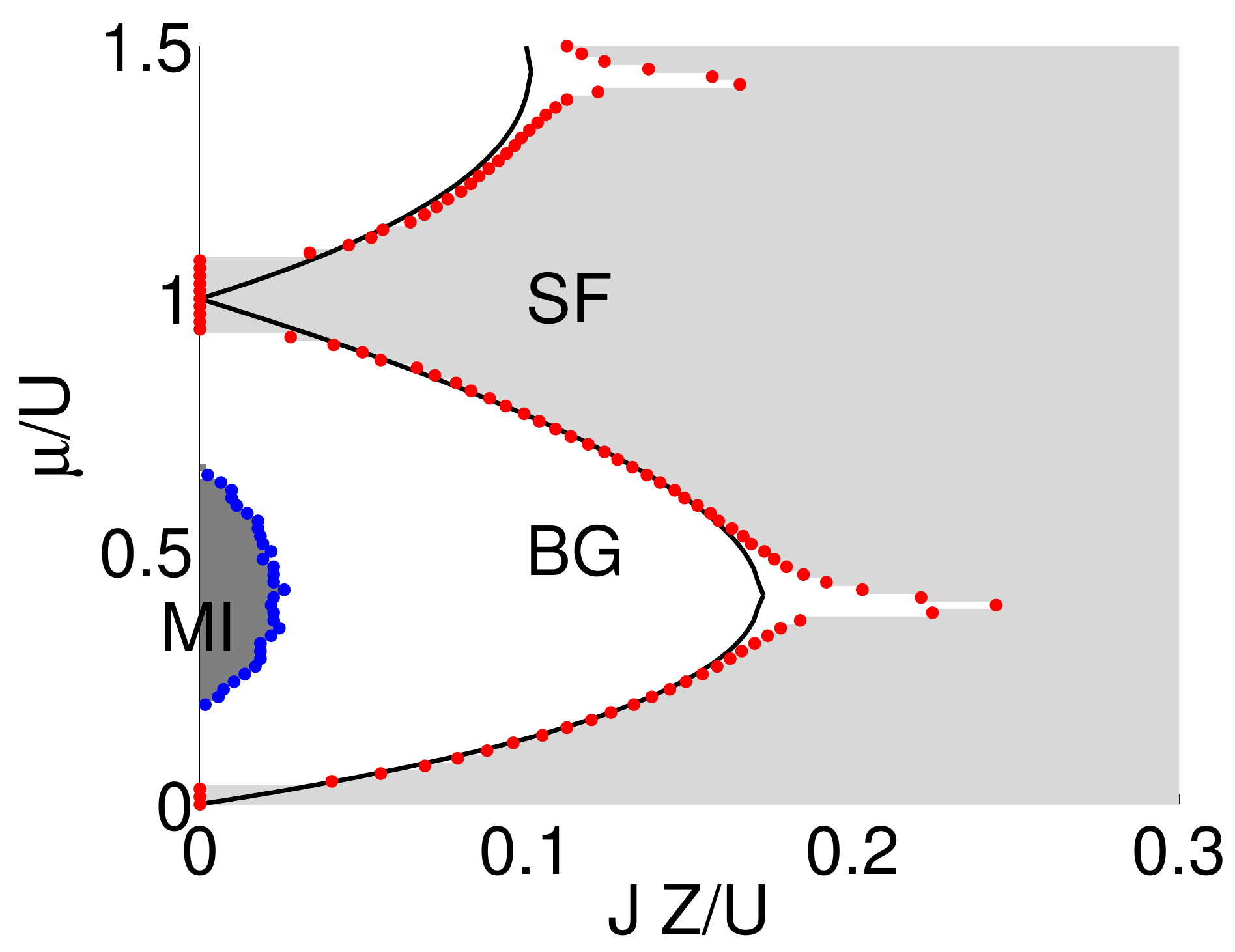}
c)\includegraphics[width=3.9cm]{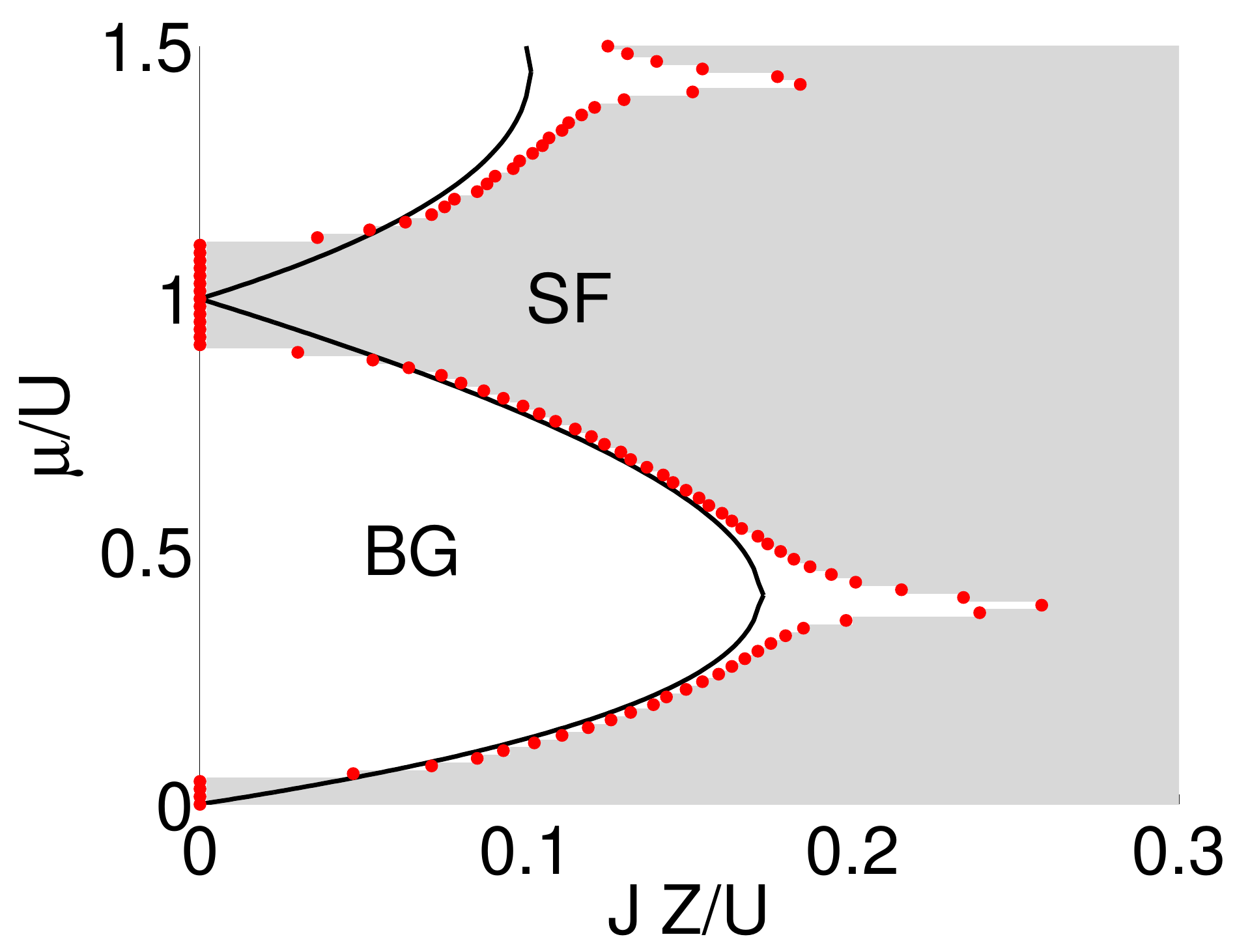}
\caption{\label{PhasenJ} Phase diagram for box distributed disordered tunneling rates $J_i$ with the disorder strength $\Delta_J/U=0.035,\,0.06,\,0.095$. The black line is the perturbative result for MI-SF transition in the ordered system given by \eqref{Oosten}.}
\end{figure}
In Figure \ref{PhasenJ} the phase diagrams resulting from LMF-Cluster-Analysis in dependence of the chemical potential $\mu/U$ and the mean tunneling rate $J Z/U$ are shown. It is important to notice that here the width of the disorder distribution is one order of magnitude smaller than for on-site disorder. Here two new unique features occur in the phase diagram: Firstly, the BG regions are separated into individual regions by SF regions reaching down to $JZ/U\rightarrow0$. Secondly, the distance between the Mott-lobes increases with their number $n$ and the number of Mott-lobes is finite. For an intuitive explanation, we recall the perturbative result of the ordered case, given in equation \eqref{Oosten}, which is the limiting result for vanishing disorder strength $\Delta_J\rightarrow0$. For a small but non-zero disorder strength $\Delta_J$, the first SF regions in the lattice occur at sites with $J+\Delta_J/2$. Consequently, the boundary of the MI region is shifted to the left by $-\Delta_J/2$ with respect to ordered case with tunneling rate $J$. Since in the ordered case the tip of the Mott lobes (at tunneling rate $J_{\rm max}(n)$) decreases with increasing $\mu$ as $1/\mu$ the shifted Mott lobes for tunneling disorder strength $\Delta_J$ must disappear when $1/\mu$ gets smaller than $\Delta_J$ resulting in only a finite number of Mott lobes. With other words: Even at a vanishing average tunneling rate ($J=0$) a non-vanishing disorder in the tunneling rate produces an increasing fraction of SF sites when $\mu$ is sufficiently large.\\
To put it quantitatively: Along the $\mu/U$-axis ($J=0$) the MI-lobes exist between $\mu_-\left(\Delta_J/2,Z,U,n\right)$ and $\mu_+\left(\Delta_J/2,Z,U,n\right)$, with $\mu_\pm$ given by equation \eqref{Oosten}. For fixed disorder strength $\Delta_J$ the height of the Mott-lobes is given by \mbox{$\Delta_{\mu}^{\text{MI}}=2 \sqrt{\frac{1}{4}\left(\frac{\Delta_J Z}{2 U}-1\right)^2-\frac{\Delta_J Z}{2 U} n}$}, which decreases with $n$. The width becomes zero for \mbox{$n_c^{\text{MI}}=\frac{1}{2} \left(\frac{\Delta_J Z}{2 U}-1\right)^2\frac{U}{\Delta_J Z}$}, which means that only a finite number $n_c^{\rm MI}$ of Mott lobes exist. As a consequence, the Mott-lobes disappear one after the other for increasing disorder strength $\Delta_J$. The last Mott-lobe ($n=1$) disappears at \mbox{$\frac{\Delta_J}{U}=\frac{3-2\sqrt{2}}{2}\approx0.0858$}. This is different from the on-site disorder case, where all vanish at the same critical disorder strength.  In \mbox{Figure \ref{PhasenJ} a)} \mbox{($\Delta_J/U=0.035$)} three Mott-lobes exist, two of which are visible, while in b) \mbox{($\Delta_J/U=0.06$)} only one remains. In the last diagram \mbox{($\Delta_J/U=0.095$)} no Mott-lobe exists, as the critical disorder strength is exceeded.\\
The Mott-lobes are surrounded by the BG phase. As a new feature in comparison to on-site disorder case we find disconnected BG regions between $\mu_-\left(\Delta_J/2,Z=1,U,n\right)$ and $\mu_+\left(\Delta_J/2,Z=1,U,n\right)$, with $\mu_\pm$ given by equation \eqref{Oosten}, which are separated from each other by the SF region in the vicinity of integer values of $\mu/U$. The fact that the SF region survives in the limit $J\rightarrow0$, is a unique feature of tunneling disorder. The width of the BG regions along the $\mu/U$-axis is given by \mbox{$\Delta_{\mu}^{\text{BG}}=2 \sqrt{\frac{1}{4}\left(\frac{\Delta_J}{2 U}-1\right)^2-\frac{\Delta_J}{2 U} n}$}, which also decreases for growing $n$. The number of BG regions is given by \mbox{$n_c^{\text{BG}}=\frac{1}{2} \left(\frac{\Delta_J}{2 U}-1\right)^2\frac{U}{\Delta_J}$}. Even though the BG regions survive for even higher disorder strength than the MI-lobes, they analogously disappear one after the other and finally disappear completely at \mbox{$\frac{\Delta_J}{U}=2\left(3-2\sqrt{2}\right)\approx0.3431$}. The SF phase exists for infinitesimal small tunneling rates between these BG regions. At the ends of the BG regions narrowing tips occur, which are located along the line of mean integer filling but finally end in the SF region. 

\subsection{Disorder in $U$}\label{section:DissU}
\begin{figure}[!htb]
a)\includegraphics[width=3.9cm]{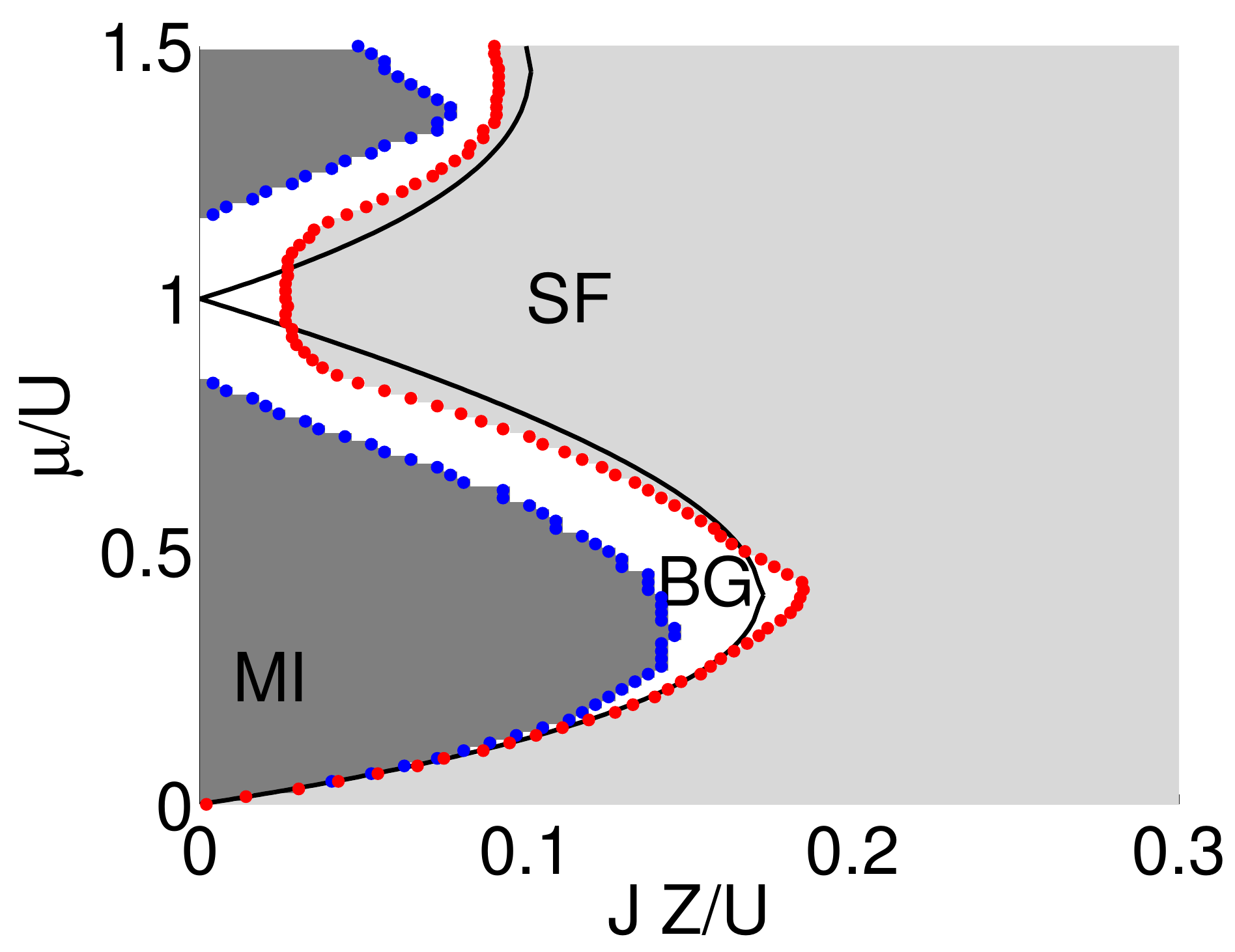}
b)\includegraphics[width=3.9cm]{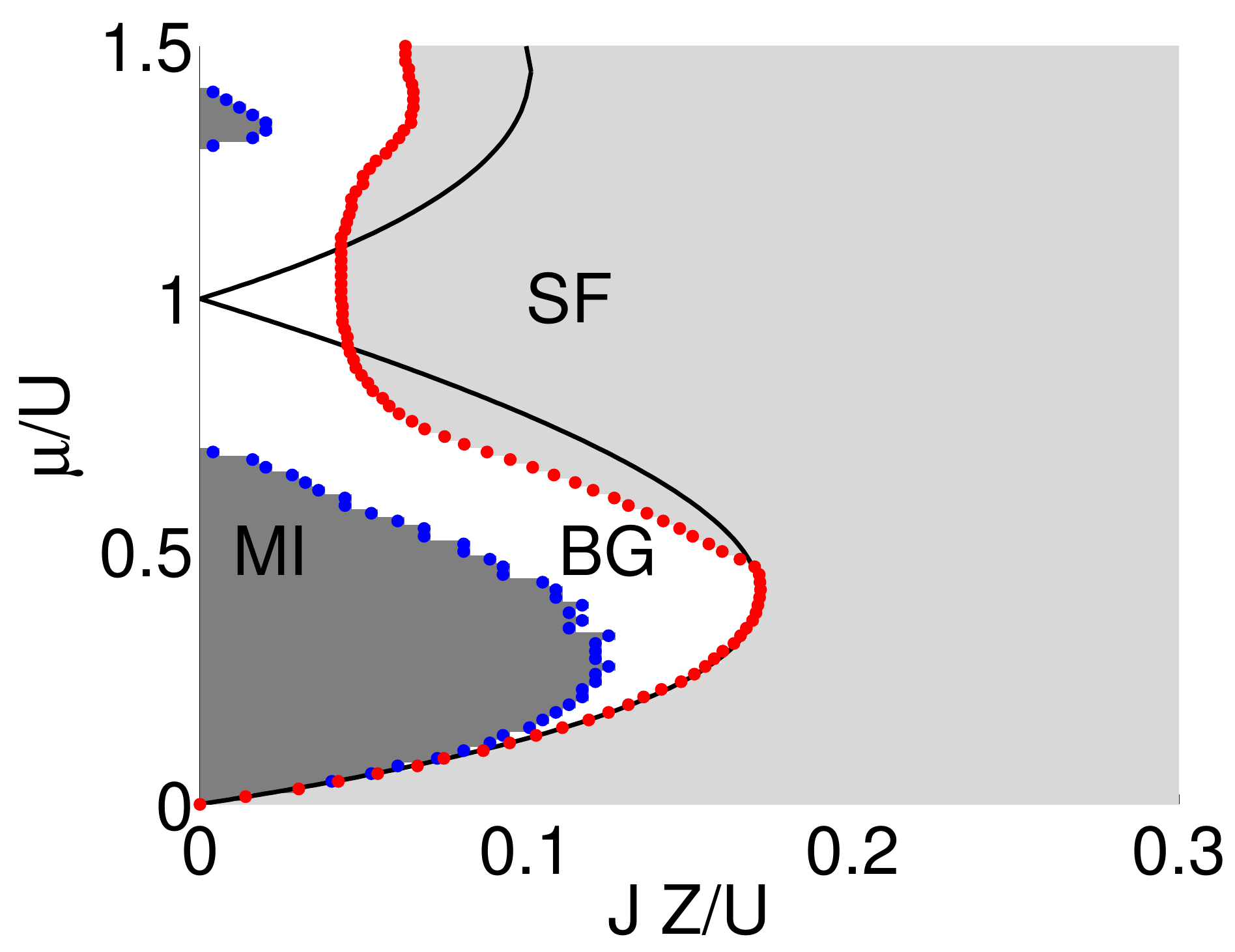}
c)\includegraphics[width=3.9cm]{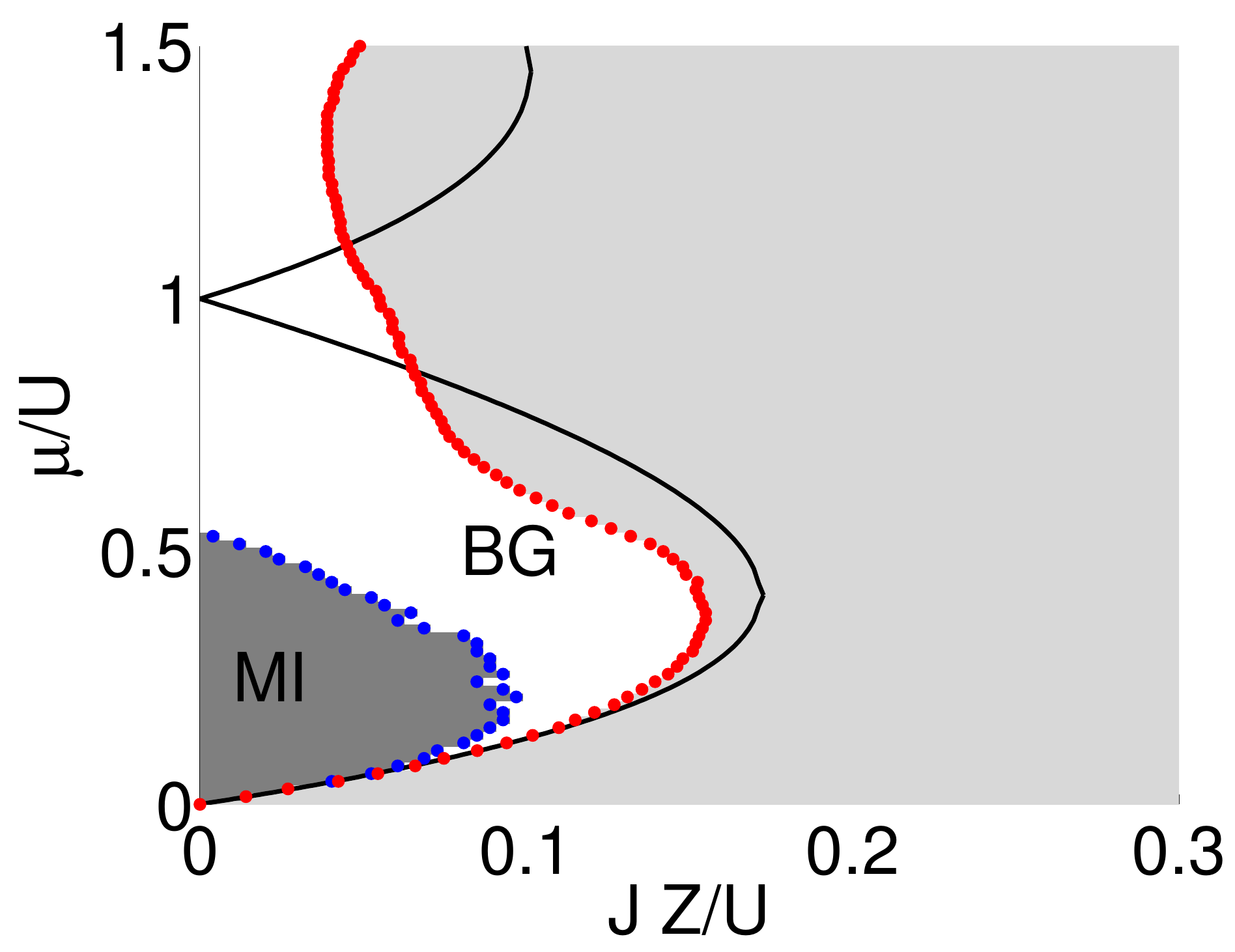}
d)\includegraphics[width=3.9cm]{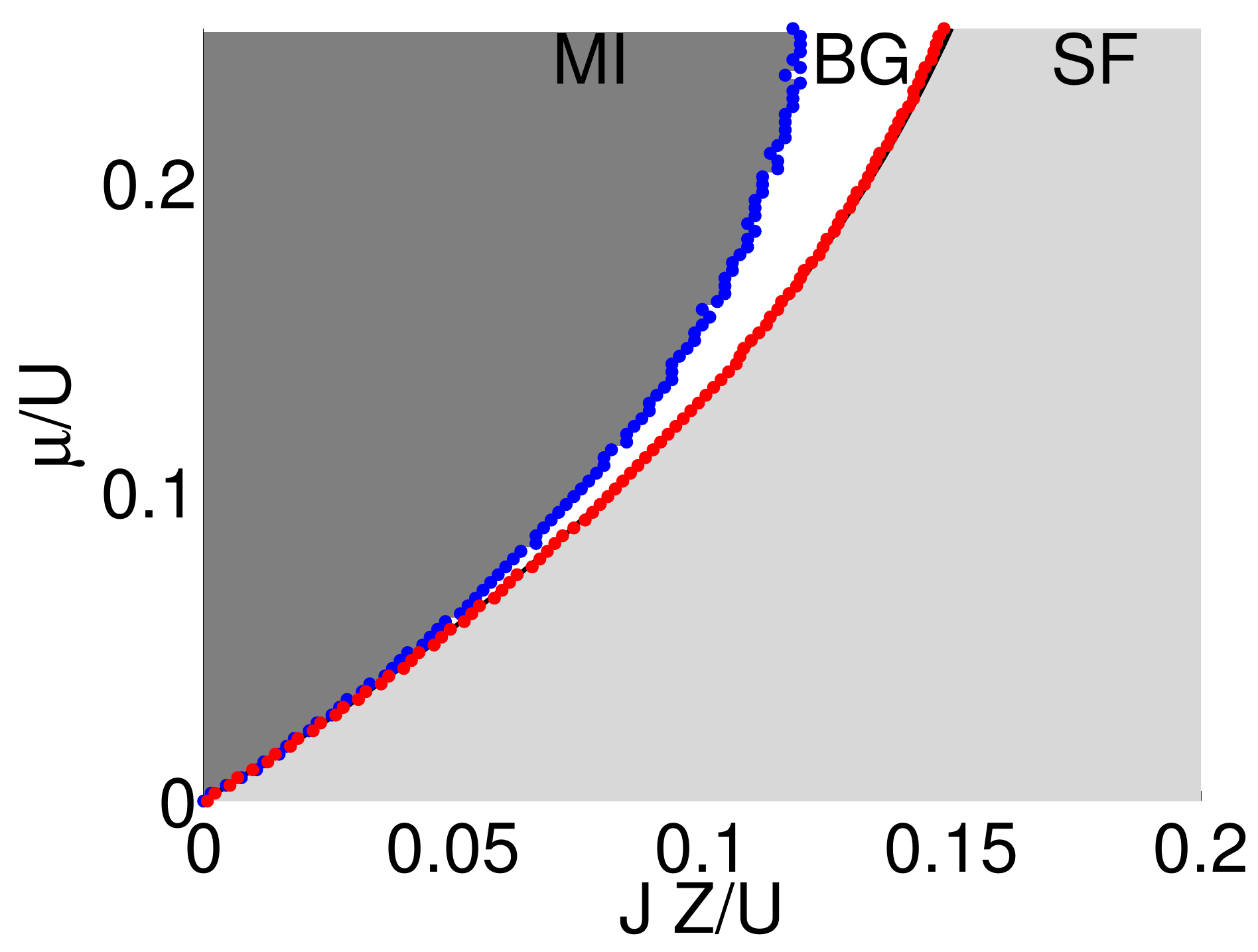}
\caption{\label{PhasenU} Phase diagram for box distributed disordered two-particle interactions $U_i$ with the disorder strength \mbox{$\Delta/U=0.35,\,0.6,\,0.95$} and a blow-up for $\Delta=0.6$ showing the tricritical point at $\mu/U=J Z/U=0$. The black line is the perturbative result for MI-SF transition in the ordered system given by \eqref{Oosten}.}
\end{figure}

Disorder in the inter-particle interaction can be realized near the Feshbach resonance \cite{Gimp05,Wild05} and a uniform distribution of this parameter has been studied in \cite{Gimp05,Gimp06}. The phase diagrams for this case resulting from LMF cluster analysis are shown in Figure \ref{PhasenU} for increasing disorder strength, where $U$ is the mean value of the disorder distribution $p(U_i)=U+\Theta\left(\Delta_U/2-|U_i|\right)/\Delta_U$. Analogously to the disordered tunneling case we find a finite number of Mott-lobes. 
Intuitively this can be understood by recalling the MI boundaries \eqref{Oosten} of the ordered case, as we have already discussed for tunneling disorder. For small tunneling rates, the first SF sites occur, where the tunneling rate $J$ overcomes the reduced inter-particle interaction $U-\Delta_U/2$. Thus, the Mott-lobes shrink all by the same amount for fixed disorder strength $\Delta_U$. Therefore, the smallest Mott-lobes of the ordered system disappear leading to a finite number of Mott-lobes. Along the $\mu/U-$axis they extend from $\mu_-=\left(n-1\right)\left(U+\frac{\Delta_U}{2}\right)$ to $\mu_+=n\left(U-\frac{\Delta_U}{2}\right)$ and they disappear at a critical disorder strength $\Delta_U^c/U=\frac{2}{2 n-1}$, where $\mu_-$ and $\mu_+$ meet \cite{Gimp05}. As the critical disorder strength $\Delta_U^c$ depends on the number $n$ of the specific Mott-lobe, they vanish one after another, until for $\Delta_U^c/U=2$ the first Mott-lobe is the last to disappear. For all disorder strengths there is only one connected BG region, respectively one SF region. This is different from the system with tunneling disorder, but analogous to the on-site disordered case.\\
A new feature occurs below the first Mott-lobe, which is shown as a blow-up in Figure \ref{PhasenU} d): In this region the BG-SF transition widely follows the MI-SF transition of the ordered case. Between this transition line and the MI-BG transition line the BG 
phase forms a narrowing strip. Both transition lines approach each other tightly for small tunneling rates and form a tricritical point in the limit $\mu/U\rightarrow0$ and $J Z/U\rightarrow0$, which does not contradict the fact that a direct transition from MI to SF is impossible in the disordered case \cite{Gura09, Poll09}.\\
This phenomenon can be understood recalling \mbox{equation \eqref{Oosten}} and studying the behavior of this equation under a variation of $U$: From its derivation, \mbox{equation \eqref{Oosten}} describes the transition line, at which the SF order parameter $\psi$ becomes non-zero in the ordered case. In the disordered case this takes place at the BG-SF transition. Therefore, the BG-SF transition line in the disordered case follows equation \eqref{Oosten} in regions, where it is stable against variation of $U$. This variation of $\mu_\pm\left(J,Z,U,n\right)$, which results from perturbation theory for small $J$ \cite{Oost01}, under a change of $U$ is given by the derivative $\frac{\partial \mu_\pm}{\partial U}$. The expansion of the derivative for small $J$ is given by
\begin{equation}
 \frac{\partial \mu_\pm}{\partial U}\approx\begin{cases}x^2 n \left( n+1\right)+n&\text{upper branch,}\\
 -x^2 n \left( n+1\right)+n-1&\text{lower branch,}\end{cases}
\end{equation}
 in powers of $x=\frac{J Z}{U}$. Notice that the linear term cancels and in general is different from zero. Only in case of the lower branch of the first Mott-lobe ($n=1$) it vanishes for zero tunneling rate. For increasing tunneling rates it grows less then linearly, since $x$ is smaller than one. This means that the lower branch of the first Mott-lobe is fairly stable against variation of $U$. For all other Mott-lobes $n>1$ the absolute value of the derivative is positive for small tunneling rates $J$. This feature of the lower branch of the first Mott-lobe is unique and does not occur for other branches of the disordered inter-particle interaction case. Therefore, the BG-SF transition of the inter particle interaction disordered system below the first Mott-lob widely follows the transition line of the ordered system, which is given by equation \eqref{Oosten}, leading to the tricritical point at the origin of the phase diagram.
 
\section{Bichromatic potential}\label{quasi-periodic}
Experimentally disorder can be introduced either by a diffuser \cite{Whit09} or by a bichromatic potential \cite{Lye05,Fall07,Whit09}. The diffuser modifies the intensity of the laser, which leads to inhomogeneities in the resulting optical lattice. For a detailed comparison with theoretical predictions a thorough characterization of the diffuser is necessary. Especially the width of the disorder distribution is a crucial system parameter, which is fixed by the diffuser and cannot be tuned freely. Alternatively a quasi-periodic potential is formed by a main optical lattices with a high intensity, which is superposed by a second weaker one with slightly different wave length \cite{Lye05}. The resulting lattice is not periodic but quasi-periodic and displays local inhomogeneities. Such a  quasi-periodic potential is the basis for our calculation, from which we will extract all BH parameters and finally discuss the resulting phase diagram in dependence of the laser intensities of both lattices for integer filling.\\
The quasi-periodic potential in two dimensions is given by
\begin{eqnarray}\label{Potential}
 V \left( x,z \right) &=& V_1 \left( \cos^2\left(k_1 x\right)+\cos^2\left(k_1 z\right)\right)\nonumber \\
 &\phantom{=}&+V_2 \left( \cos^2\left(k_2 x\right)+\cos^2\left(k_2 z\right)\right), 
\end{eqnarray}
where $k_i=\frac{2 \pi}{\lambda_i}$ ($i=1,2$), the lattice constant $a=\frac{\pi}{k_1}$ and the intensities $V_i=s_i E_{Ri}$ are given in units of the recoil energy $E_{Ri}=\frac{\hbar^2 k_i^2}{2m}$. The wavelengths are chosen to be $\lambda_1=830$ nm and $\lambda_1=1076$ nm with reference to the experimental setup of \cite{Fall07}. In experiments $\phantom{}^{87}\text{Rb}$, which has a mass of $m=1.443 10^{-25}$ kg, is widely used. The amplitude of the main lattice $s_1$ determines the depth of the lattice. The amplitude of the second lattice \mbox{$s_2\ll s_1$} is by far smaller than the first one and increases the influence of the disorder strength.\\
\begin{figure}[!htb]
\includegraphics[width=3.9cm]{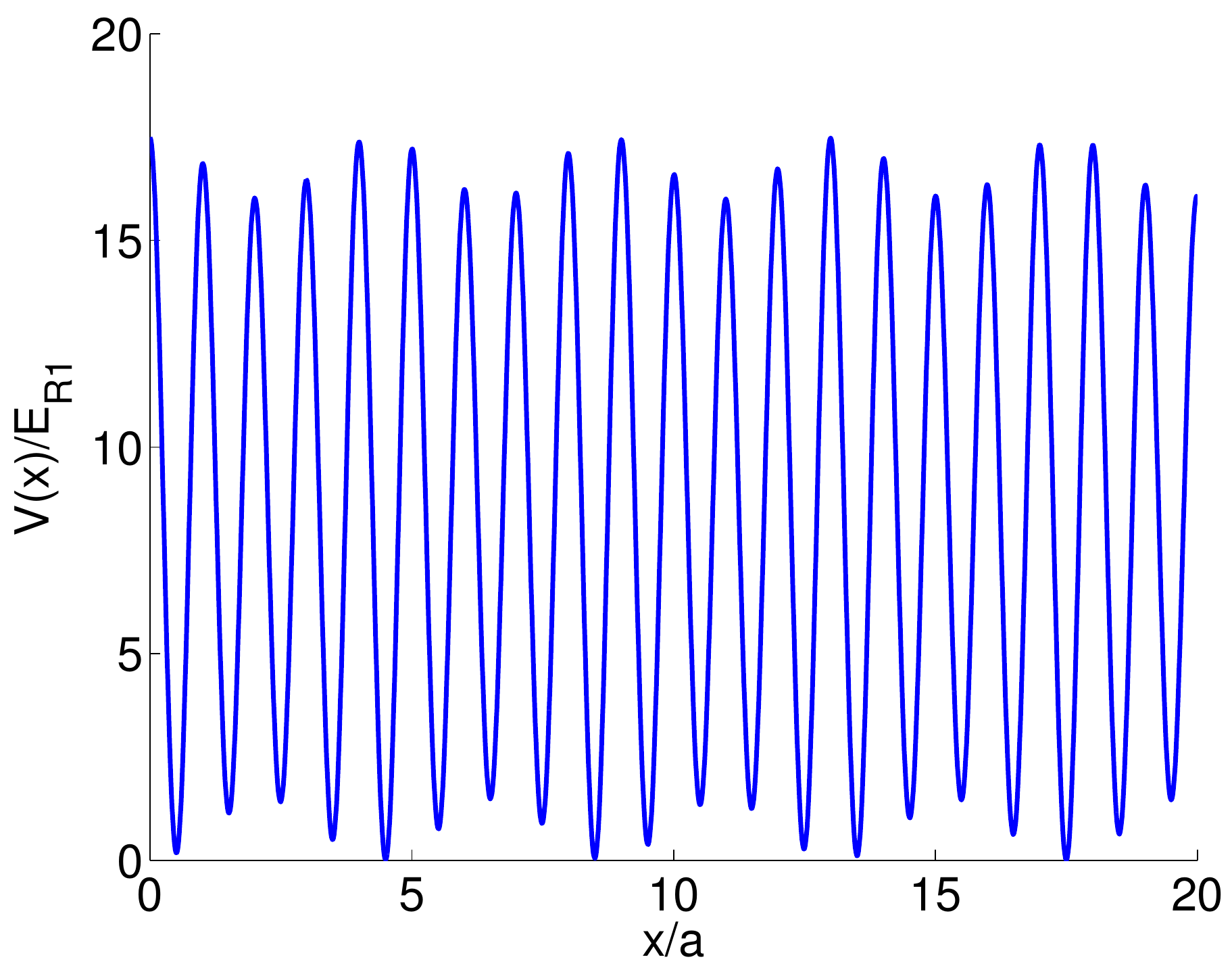}
\caption{\label{Pot} A quasi-periodic potential in one dimension according to $V \left( x \right) = V_1 \cos^2\left(k_1 x\right)+V_2 \cos^2\left(k_2 x\right)$ with $s_1=16$, $s_2=2.5$ and $V_2/V_1\approx 0.09$.}
\end{figure}

\subsection{Wannier-Functions}
\begin{figure}[!htb] 
a)\includegraphics[width=3.9cm]{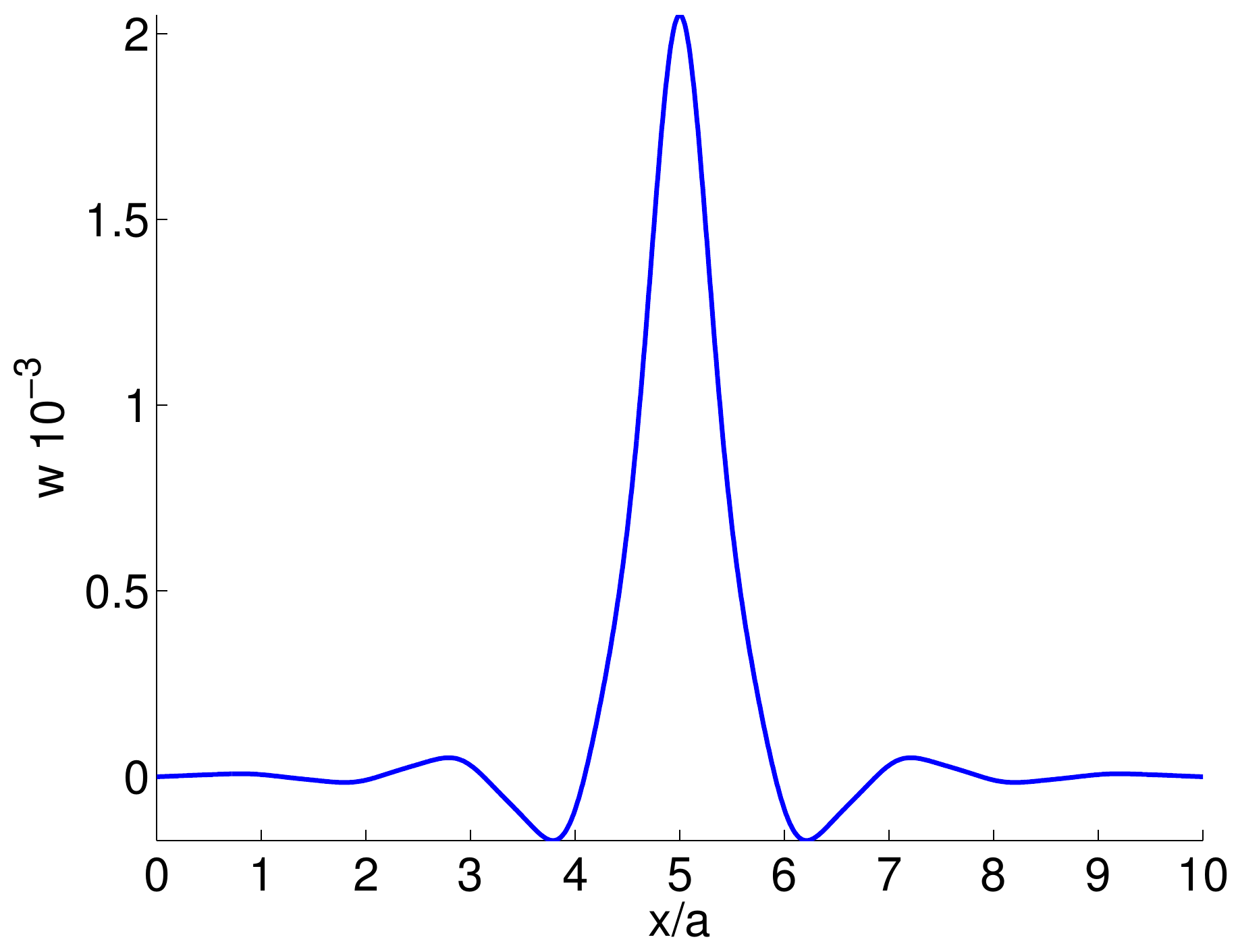}
b)\includegraphics[width=3.9cm]{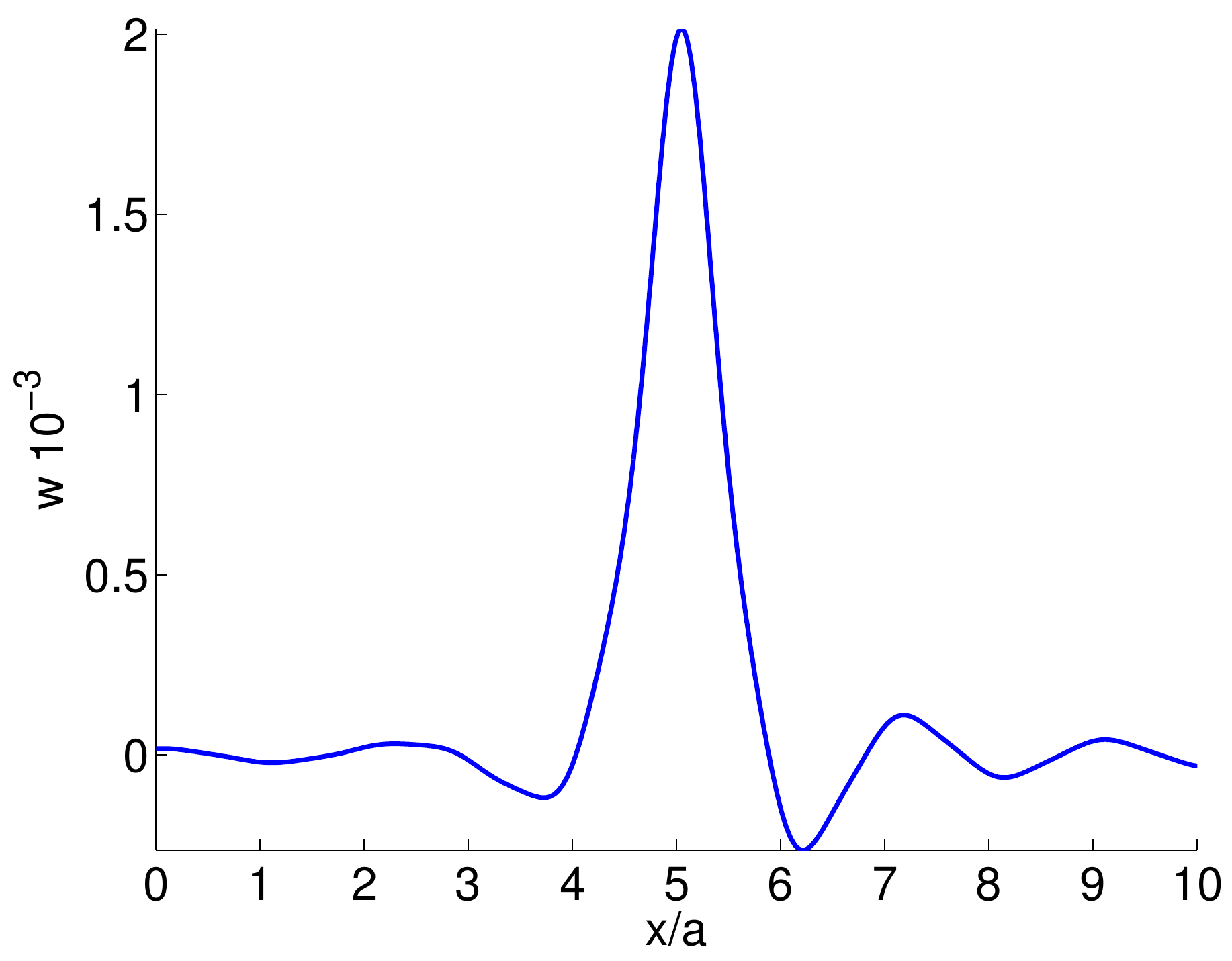}
\caption{\label{WanniePicture} Wannier functions for an ordered symmetric lattice on the left ($s_1=2$, $s_2=0$) and a disordered asymmetric lattice on the  right ($s_1=2$, $s_2=1$ with $V_2/V_1\approx 0.3$).}
\end{figure}

In order to map the quasi-periodic potential \eqref{Potential} to the BH Hamiltonian the Wannier functions for each site are computed individually. We first focus on a regular periodic $\left( V_2=0\right)$ lattice. According to the Bloch theorem the Bloch functions
\begin{equation}
 \psi_n^{\vec q}\left( \vec r \right)=u_{\vec q}\, e^{\i\, \vec q\, \vec r},\quad u_{\vec q}\left( \vec r \right)=\sum_{\vec G} c_n^{\vec q-\vec G}\, e^{-\i\, \vec G\, \vec r}
\end{equation}
solve the stationary Schrödinger equation. For every wave-vector $\vec k$ there exists a unique decomposition \mbox{$\vec k= \vec q-\vec G$}, where $\vec q$ lives in the first Brillouin zone (1.BZ). The Bloch coefficients $u_{\vec q}\left( \vec r \right)$ are periodic functions with the same periodicity as the lattice. The Bloch functions, which spread over the hole lattice, form an orthonormal basis. Thus, the Wannier functions localized at site $\vec i=\left(i_x,i_z\right)$ can be construed as
\begin{equation}\label{WannierFunc}
 W_n^i\left( \vec r \right)=\sqrt{\frac{2 \pi}{a}} \frac{1}{M} \sum_{\vec q \in 1.BZ} \psi_n^{\vec q}\left( \vec r \right)\, e^{\i\, \vec q\, \vec x_l},\quad \vec x_l=\vec l\, a.
\end{equation}
They are real functions, which fulfill \mbox{$\int_V \d V {W_n^i}^2\left( \vec r \right)=1$}. Moreover, they are symmetric due to the underlying lattice symmetry. In Figure \ref{WanniePicture} a) a Wannier function for a regular 1D lattice is shown exemplaryly. This wave function shows a dominant occupation probability at one single site and small probability at the neighboring sites.\\
In the non-symmetric case $\left(0 \neq V_2\ll V_1\right)$ the functions  
\begin{equation}\label{genWannierFunc}
 \psi_n^{\vec q}\left( \vec r \right)=\sum_{\vec G} c_n^{\vec q-\vec G}\, e^{\i\, \left(\vec q-\vec G\right)\, \vec r}
\end{equation}
still form an orthonormal basis, but their coefficients $u_{\vec q}\left( \vec r \right)$ are no longer periodic. But still, localized functions according to generalized Wannier functions can be constructed according to equation \eqref{WannierFunc}. As one can see in Figure \ref{WanniePicture} on the right, these functions are still localized at a specific lattice site, but they are asymmetric, reflecting the asymmetry of the underlying lattice.\\
With this generalization of the Wannier function for asymmetric lattice systems we follow the common derivation introducing the BH model in literature \cite{Pita03, Peth08} and thus fundamentally introduce disorder via quasi-periodic potentials. Other approaches avoid this step and describe this effect on the BH parameter effectively \cite{Guar07, Rosc08, Roux08, Deng08,Zhan14}. As we were able to show it is not sufficient to use the symmetric Wannier functions \cite{Schm09,Schm10} as an approximation in the disordered case; this is only accurate for $s_2<0.1$. Thus, in the following we will use these generalized Wannier functions in order to determine the BH parameters.   

\subsection{Bose-Hubbard Parameter}\label{Bose-Hubbard Parameter}
\subsubsection{Determination}\label{Determination}$\;$ \\
Bosons in any potential $V\left(\vec{r}\right)$ in the quasi-ideal regime are described by the quantum field theory Hamiltonian 
\begin{eqnarray}
 \hat H&=&\int \d \vec r\, \vec{\Psi}^\dagger\left(\vec{r}\right) \left(- \frac{\hbar^2 \nabla^2}{2 m}+V\left(\vec{r}\right)\right) \vec{\Psi}\left(\vec{r}\right) \nonumber \\
 &\phantom{=}&+\frac{1}{2} \int \d \vec r\,\d \vec r'\, \vec{\Psi}^\dagger\left(\vec{r}\right)\vec{\Psi}^\dagger\left(\vec{r}'\right) U \vec{\Psi}\left(\vec{r}'\right)\vec{\Psi}\left(\vec{r}\right),
\end{eqnarray}
where $U\left(\vec{r},\vec{r}'\right)$ describes the two particle interaction \cite{Pita03}. The field operator in tight-binding approximation \begin{equation}
 \vec{\Psi}\left(\vec{r}\right)=\sum_i W_i\left( \vec r \right) \O[i]{a}
\end{equation}
can be composed by the Wannier functions \mbox{$W_i\left( \vec r \right)=W_0^i\left( \vec r \right)$} of the lowest band ($n=0$) and the creation operator $\O[i]{a}$ creating particle at site \mbox{$\vec i=\left(i_x,i_z\right)$}. In the tight-binding approximation the inter-particle interaction reduces to a point-like interaction $U\left(\vec{r},\vec{r}'\right)=U_0 \delta \left(\vec r - \vec r' \right)$ \cite{Peth08}. The effective inter-particle interaction in 2d is given by 
\begin{equation}
 U_0=\frac{\hbar^2 a_s}{m} \sqrt{\frac{8 m \pi \omega_z}{\hbar}}=5.56\, 10^{-11}\,\hbar,
\end{equation}
where $a_s=5.2$ nm is the scattering length, \mbox{$m=1.443 10^{-25}$ kg} is the mass of the $\phantom{}^{87}\text{Rb}$ atoms and \mbox{$\omega_z=6 \pi$ kHz} is the frequency of the vertical confinement \cite{Berg03,Krue07,Habi13}. The BH parameters may be calculated using the ground state Wannier function \mbox{$W_i\left( \vec r \right)=W_0^i\left( \vec r \right)$} and the actual potential $V\left(\vec{r}\right)$:
\begin{eqnarray}\label{BHPara}
\epsilon_i&=&\int \d \vec r\,\, W_i\left( \vec r \right) \left(-\frac{\hbar^2 \nabla^2}{2 m}+V\left(\vec{r}\right)\right) W_i\left( \vec r \right)\nonumber\\
U_i&=&U_0\int \d \vec r\,\, {W_i}^4\left( \vec r \right)\nonumber\\
J_{ij}&=&\int \d \vec r\,\, W_j\left( \vec r \right) \left(-\frac{\hbar^2 \nabla^2}{2 m}+V\left(\vec{r}\right)\right) W_i\left( \vec r \right)
\end{eqnarray}
In contrast to the symmetric potential, all these integrals are not necessarily positive in the asymmetric case. Here the generalized Wannier functions \eqref{genWannierFunc} as well as the potential are asymmetric and rarely configurations occur, in which especially the tunneling rate is negative. Finally, the chemical potential $\mu$ as a Lagrange multiplier for the condition of a fixed particle number $N=\sum_i \langle \O[i]n \rangle=M$ of on average one particle per site is determined recursively within LMF theory.

\subsubsection{Distributions}\label{Distributions}$\;$ \\
\begin{figure}[!htb]
a)\includegraphics[height=3.5cm]{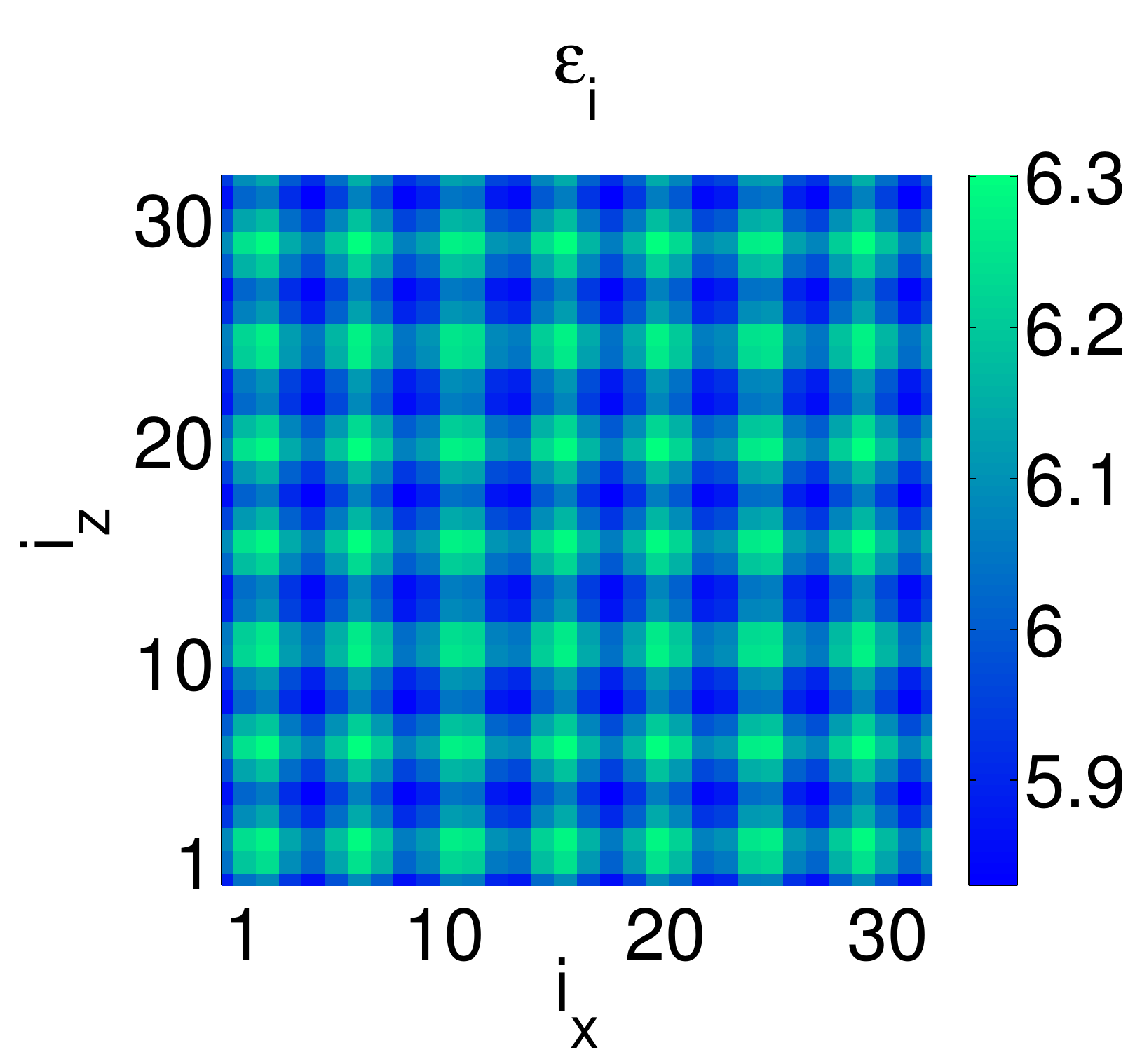}
b)\includegraphics[height=3.5cm]{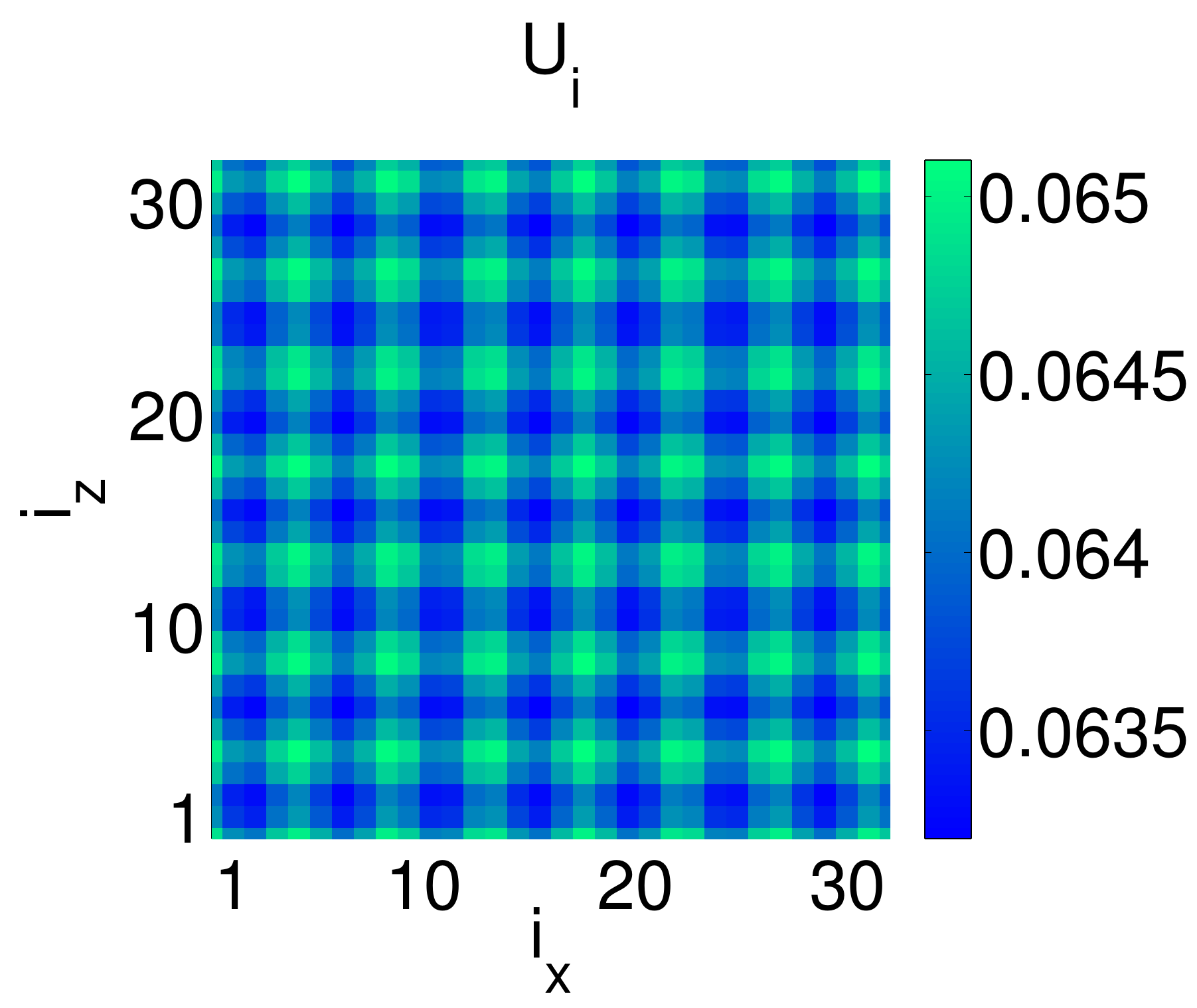}
c)\includegraphics[height=3.5cm]{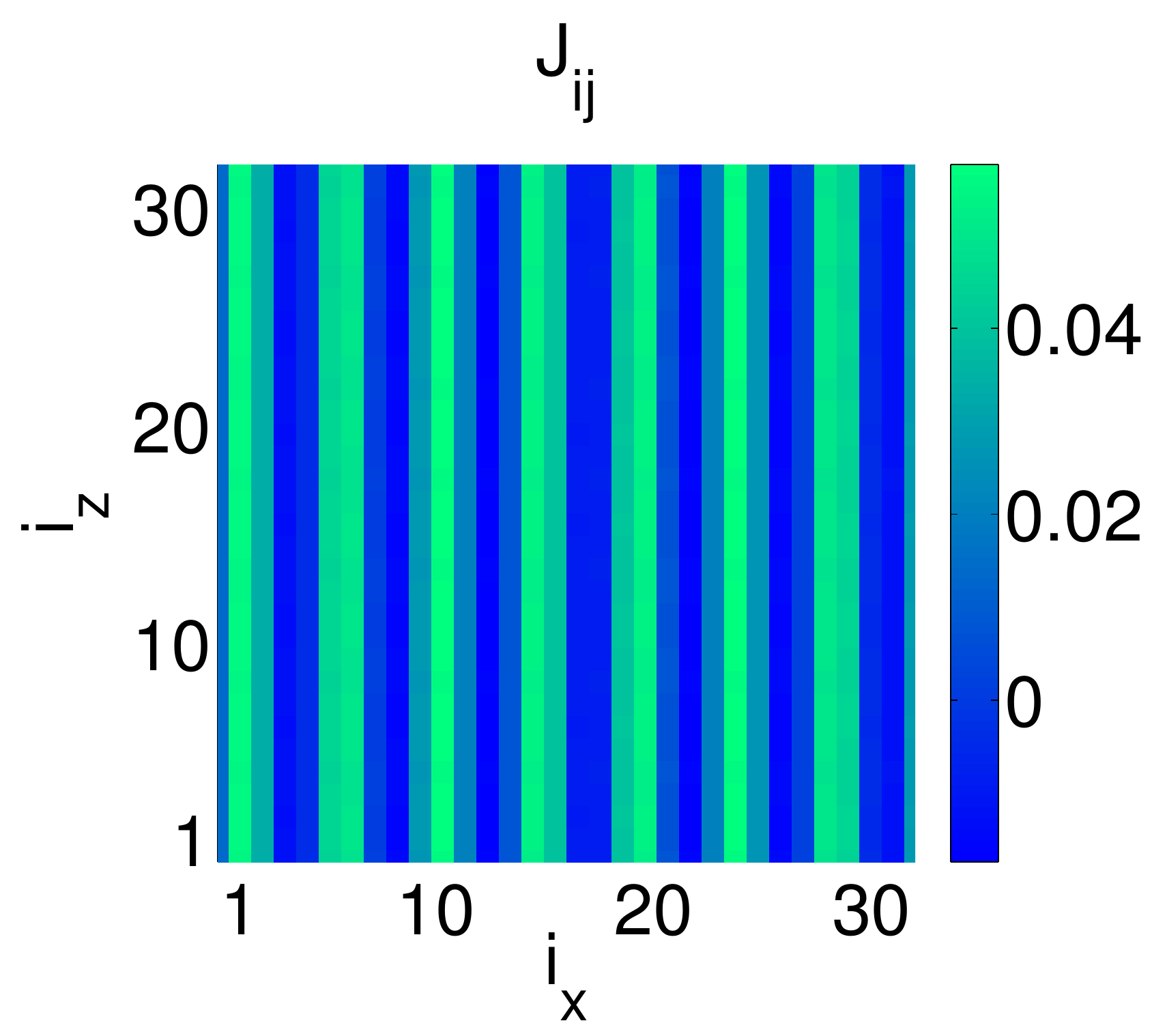}
\caption{\label{Parameter} BH Parameter for a quasi-periodic potential according to equation \eqref{Potential} with $s_1=10$ and $s_2=0.5$. The tunneling rate at site $i=\left( i_x,i_z \right)$ gives the value for the tunneling rate to the neighbor $j=\left( i_x+1,i_z \right)$. Note that this contains all information, since the tunneling rate $J_{ij}$ \eqref{BHPara} is symmetric under a change of the indices and the Potential $V \left( x,z \right)$ \eqref{Potential} under a change of the coordinates $x$ and $z$.}
\end{figure}
\begin{figure*}[!htb]
\begin{tabular}{ccc}
a)\includegraphics[width=3.9cm]{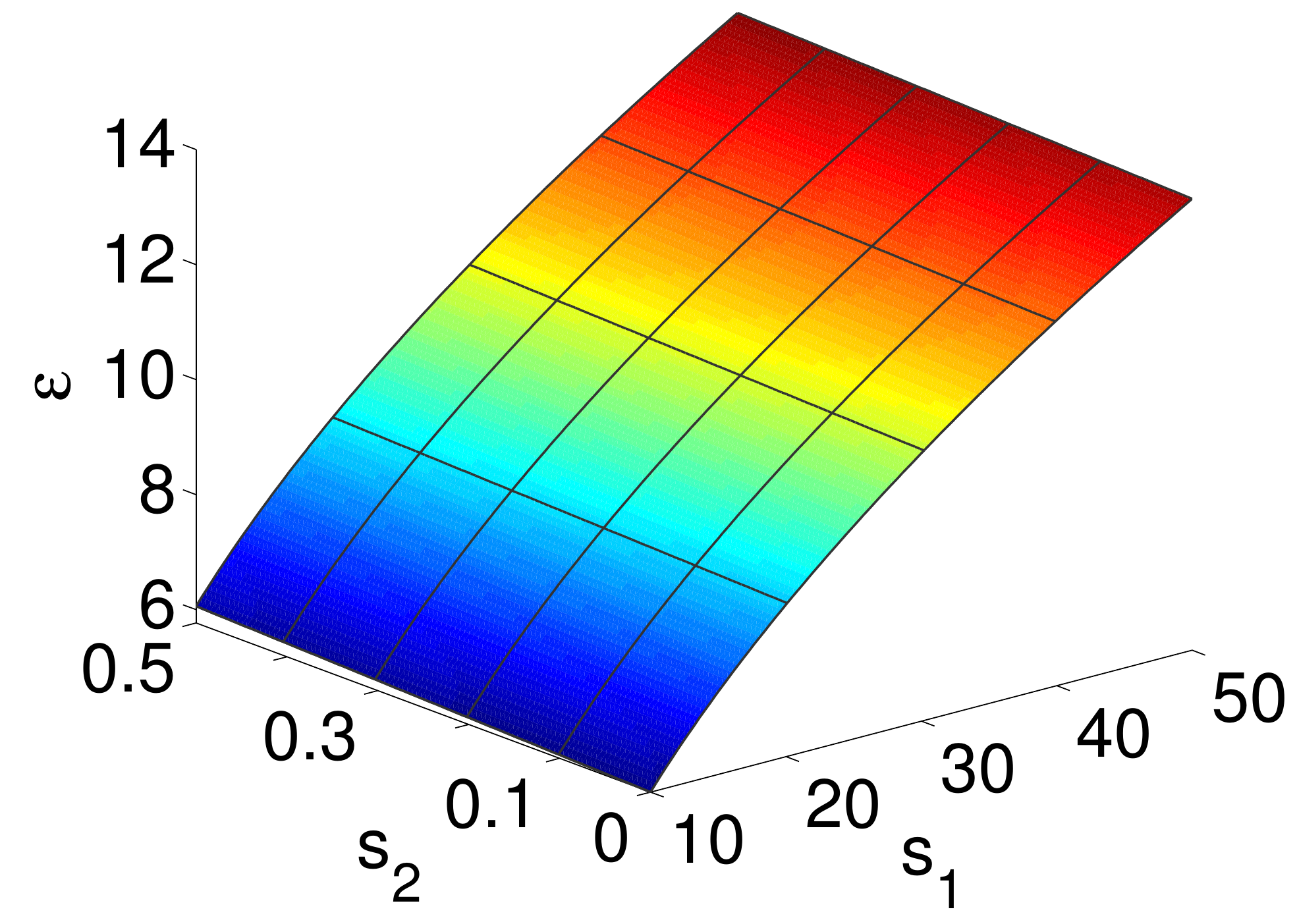} & b)\includegraphics[width=3.9cm]{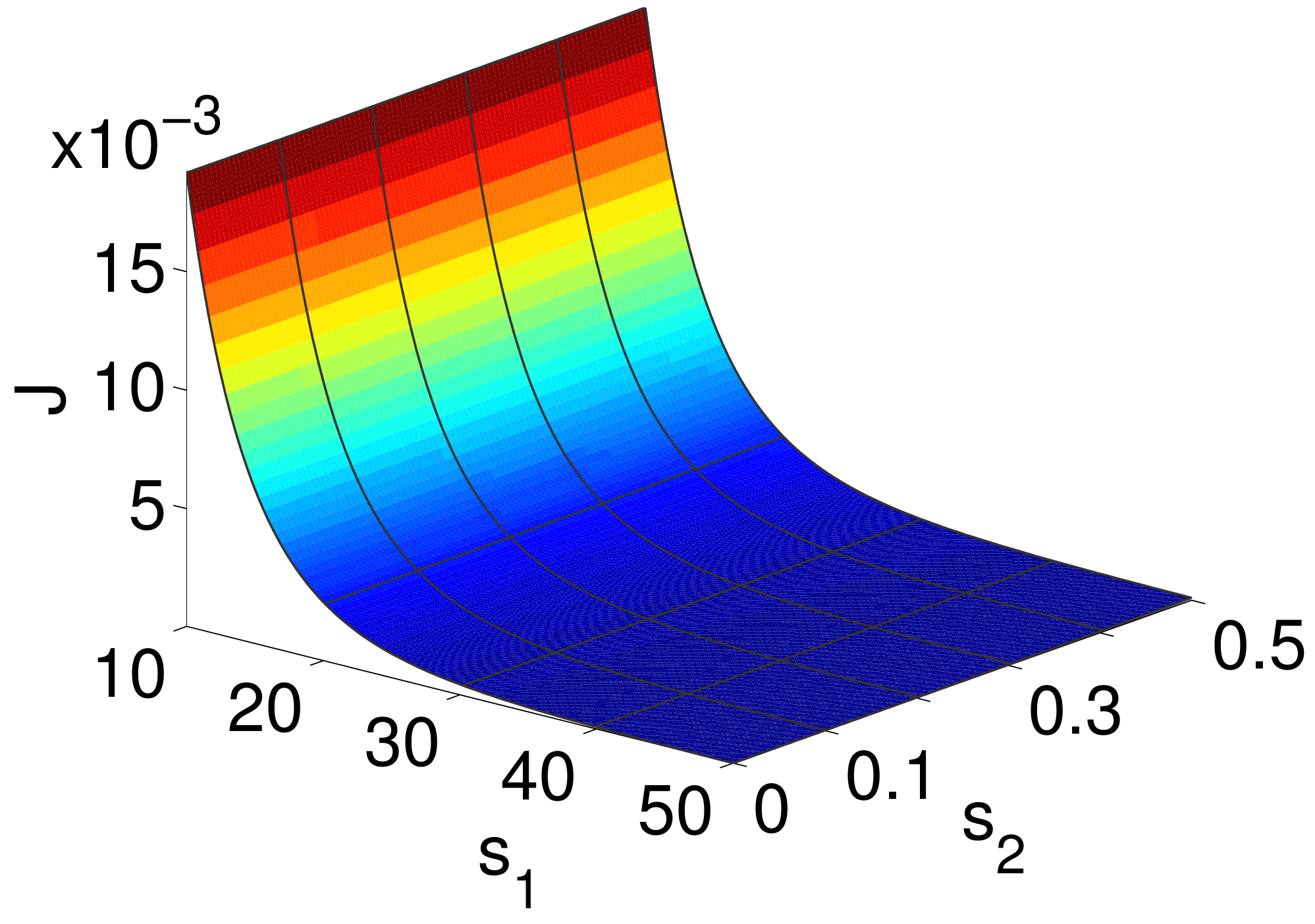} & c)\includegraphics[width=3.9cm]{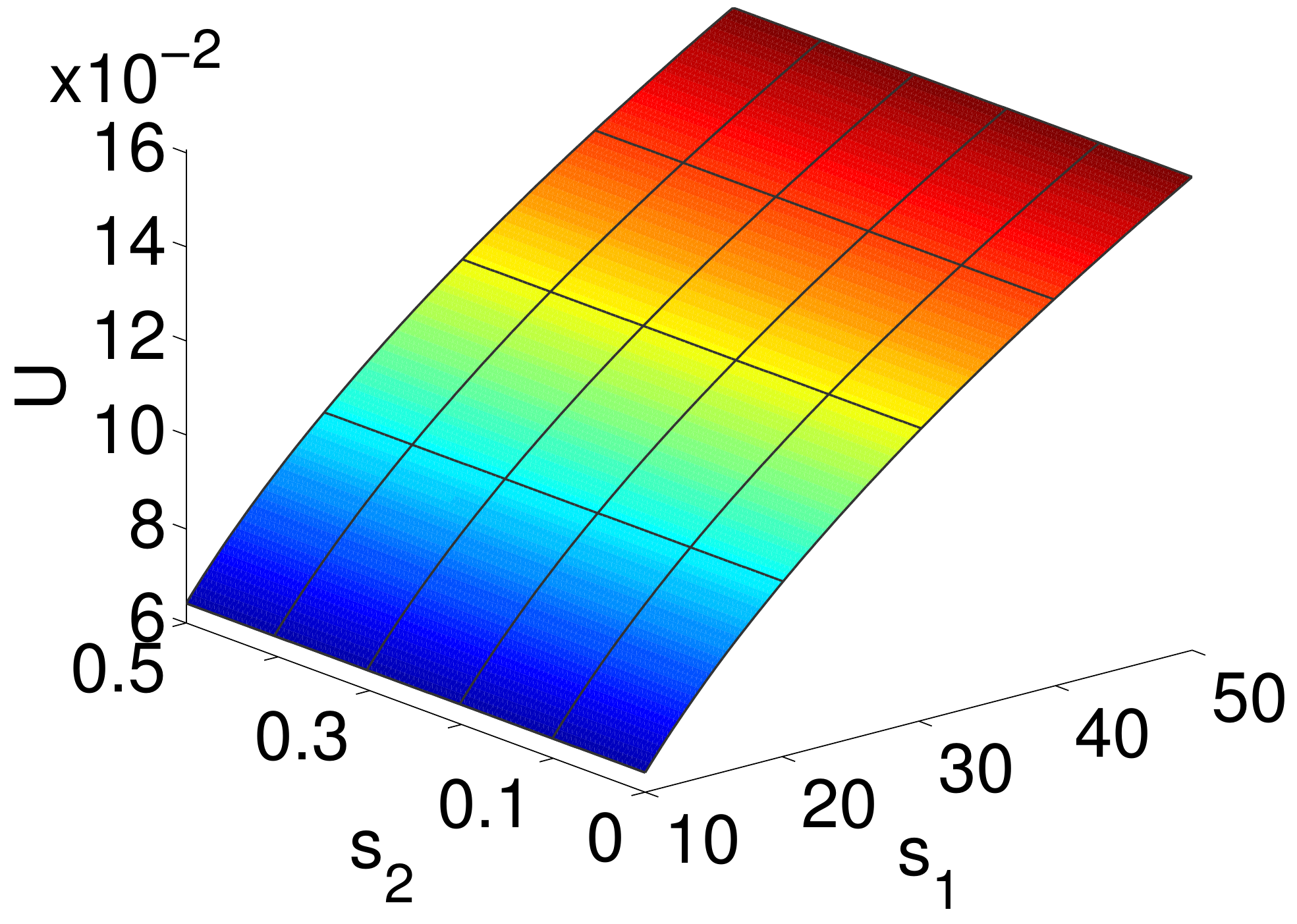}\\
d)\includegraphics[width=3.9cm]{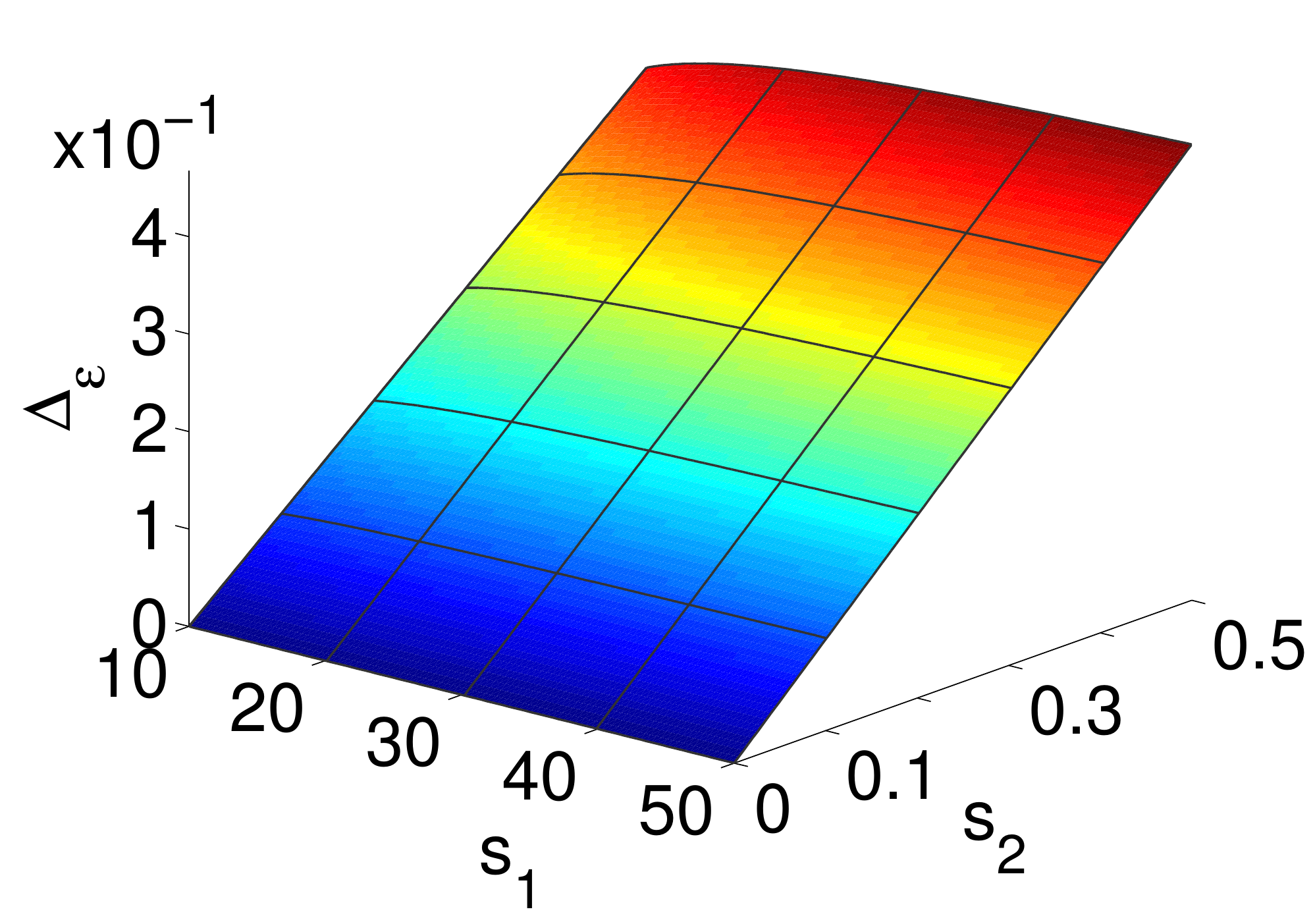} &e)\includegraphics[width=3.9cm]{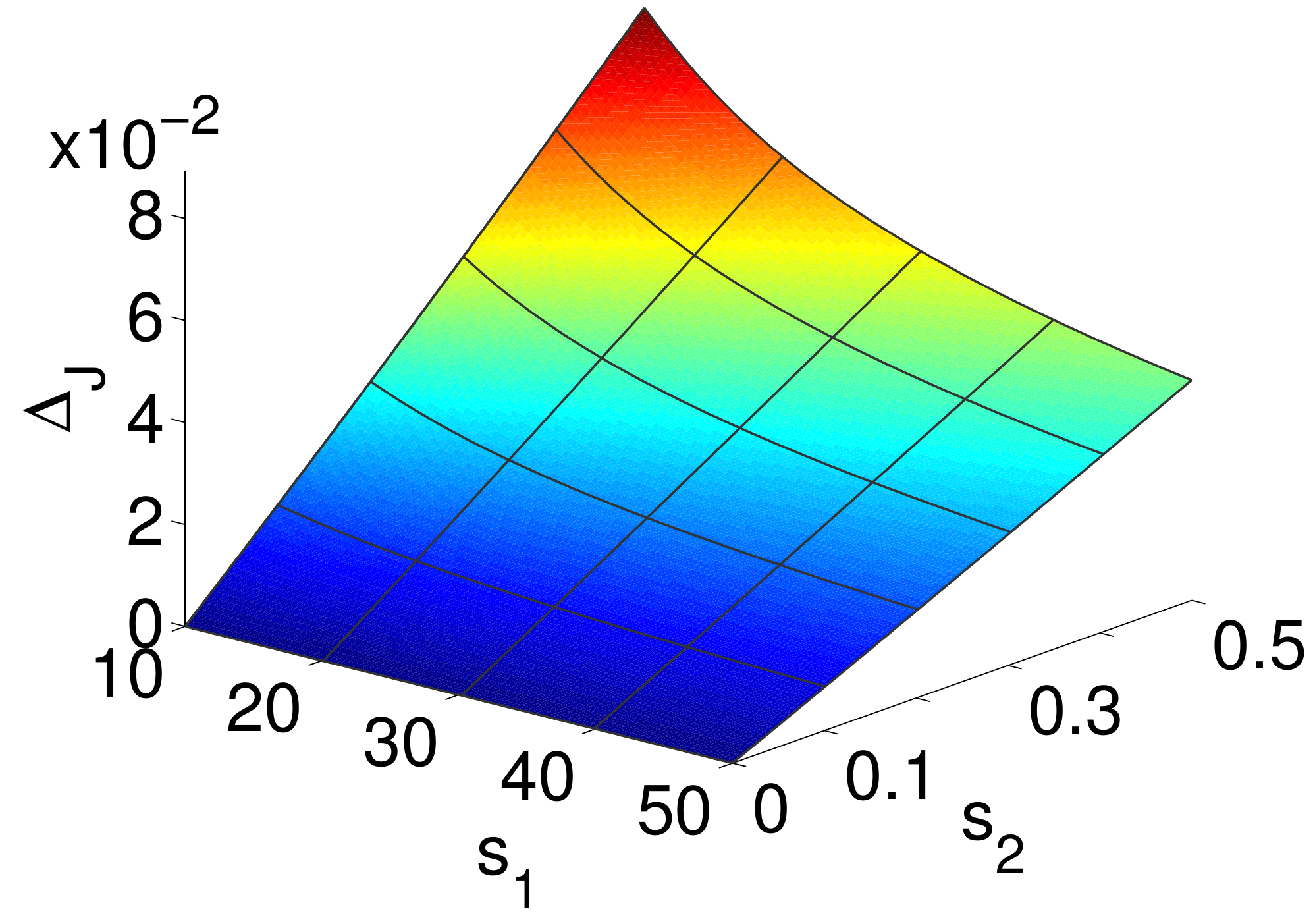} & f)\includegraphics[width=3.9cm]{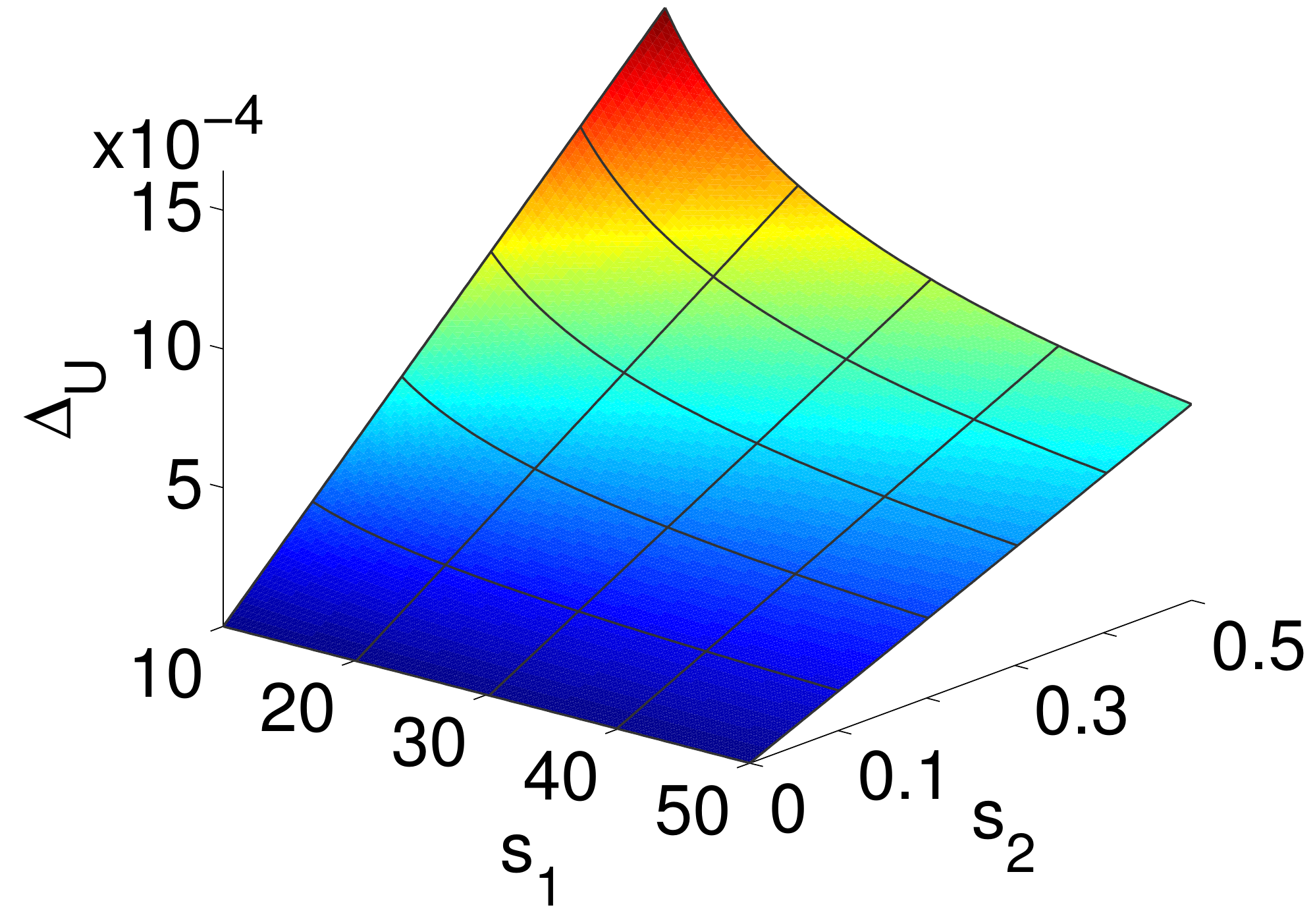}
\end{tabular}
\caption{\label{MeanValueWidth} The mean value $\alpha$ (first row) and the width $\Delta_{\alpha}=\sqrt{12 \sigma_{\alpha}^2}$ (second row) with the variance $\sigma_{\alpha}^2$ of the distributions of the BH-Parameter $\alpha=\epsilon,\,J,\,U$ in units of the recoil energy $E_{R1}$. Notice that in the figures of $\epsilon$ and $U$ the $s_1$- and $s_2$-axis are switched in comparison to the other figures.}
\end{figure*}
\begin{figure*}[!htb]
\begin{tabular}{ccc}
a)\includegraphics[width=3.9cm]{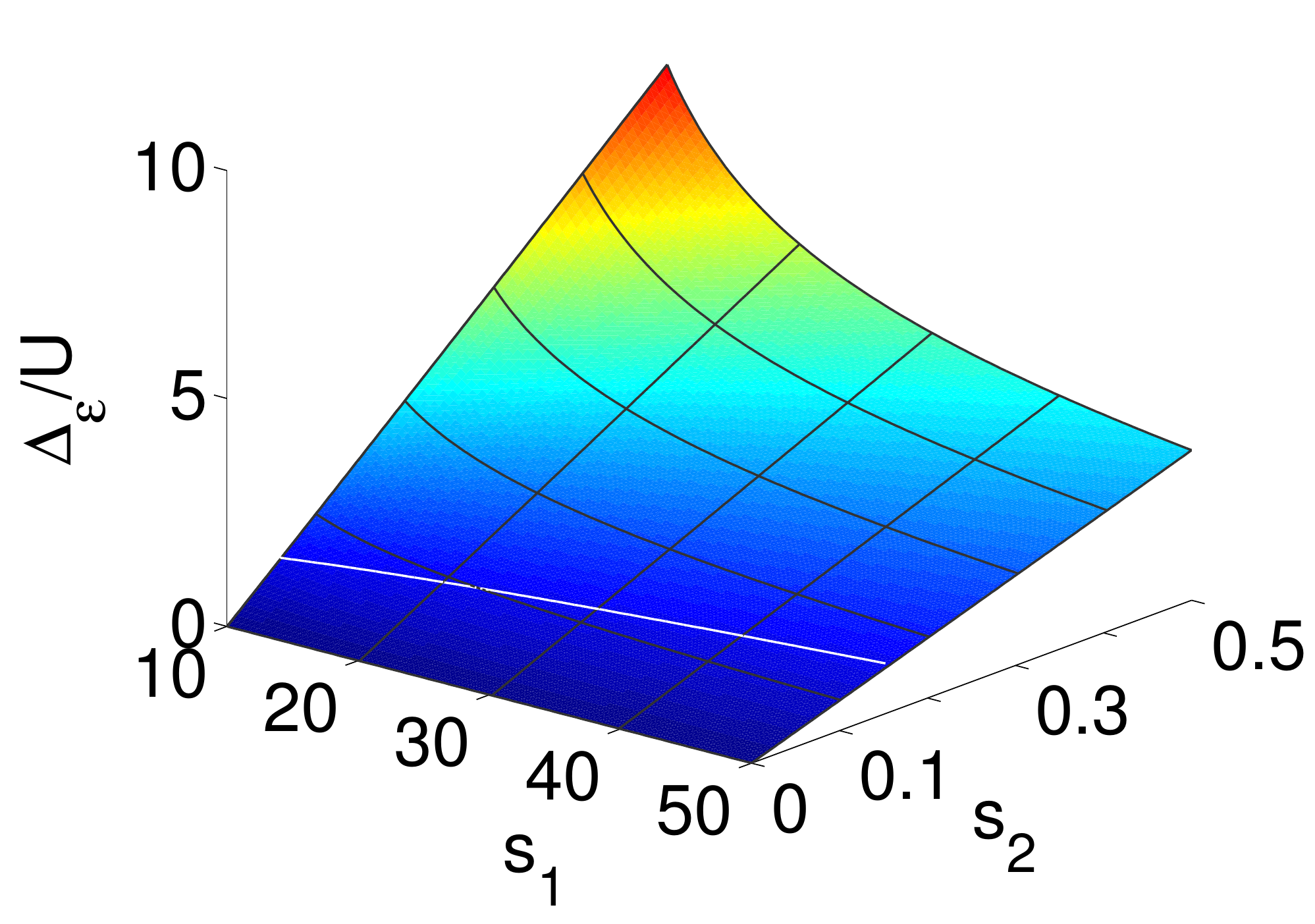} & b)\includegraphics[width=3.9cm]{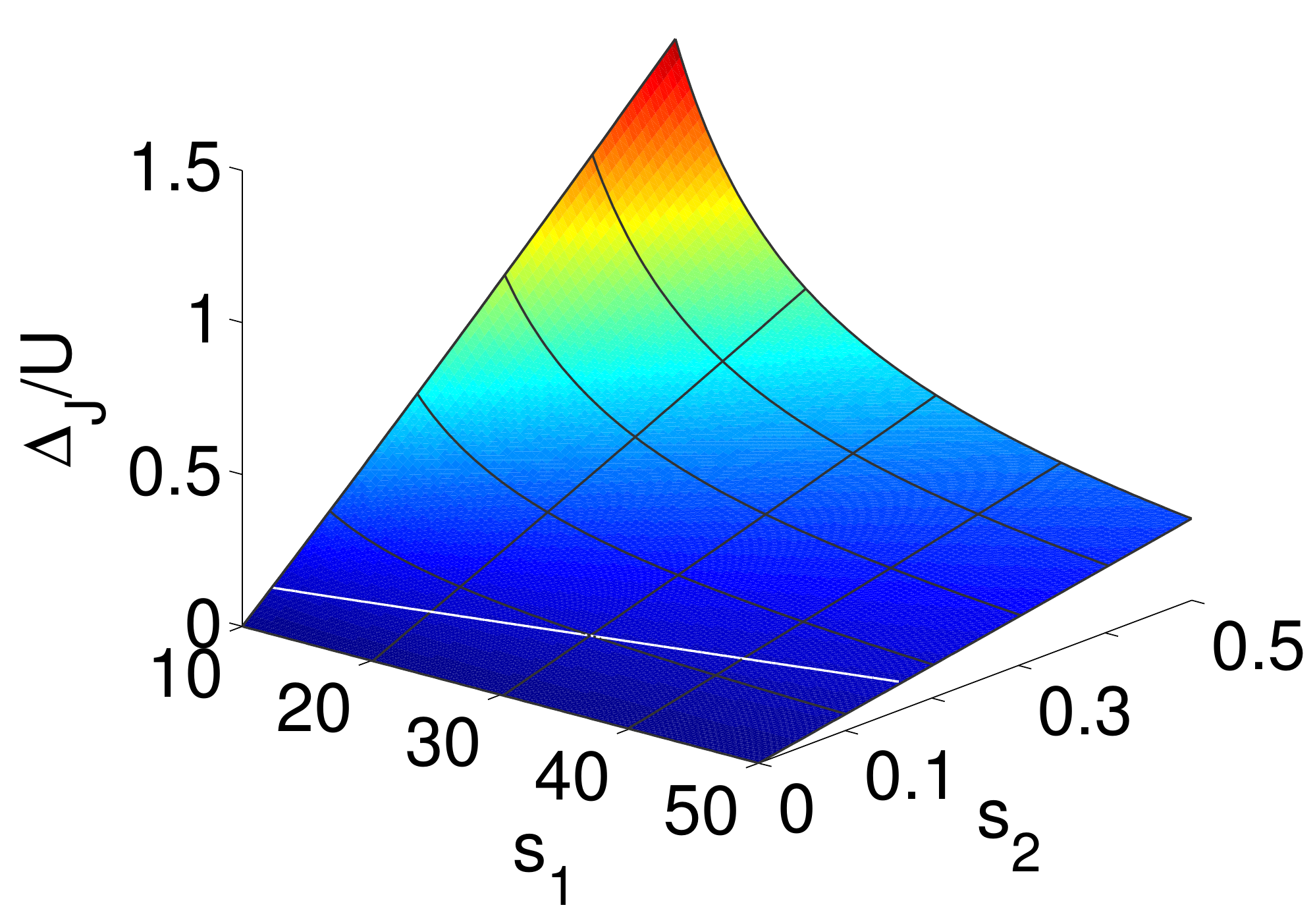} & c)\includegraphics[width=3.9cm]{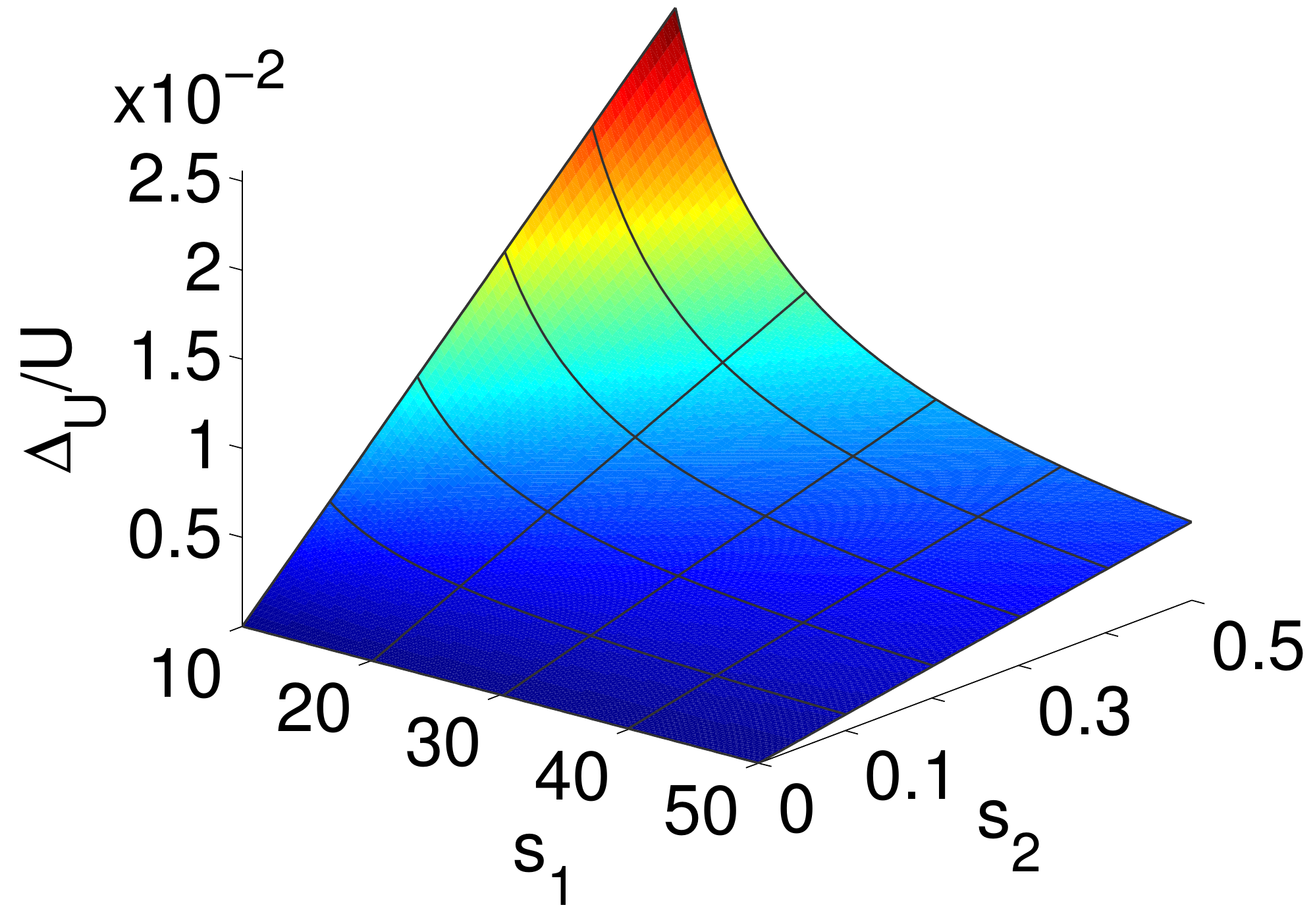}\\
$\Delta_{\epsilon}^c/U=1$& $\Delta_{J}^c/U=\frac{3-2\sqrt{2}}{2}\approx8.58\,10^{-2}$&$\Delta_{U}^c/U=2$
\end{tabular}
\caption{\label{MeanValueWidthU} The width $\Delta_{\alpha}=\sqrt{12 \sigma_{\alpha}^2}$ of the distributions of the BH-Parameter $\alpha=\epsilon,\,J,\,U$ in units $U$. The white line shows the the critical width $\Delta_{J}^c$, where the last Mott-lobe disappears, which is stated below the plot.}
\end{figure*}
The result of this procedure is a set of individual BH parameters $\alpha=\epsilon,\,J,\,U$ for each fixed parameter set $\left(s_1,s_2 \right)$. Exemplary the BH parameters for $s_1=10$ and $s_2=0.5$ are shown in Figure \ref{Parameter}. In the figures each pixel represents the value of $\epsilon_i,\,J_{ij},\,U_i$ at a specific site. The BH parameter follow the modulation of the lattice potential. Therefore, we now deal with distributions $P\left(\alpha\right)$, which depend on the parameter set $\left(s_1,s_2 \right)$ chosen for the amplitudes of the lasers. We will especially focus on their mean value $\overline{\alpha}=\int \d\alpha\,\alpha P\left(\alpha\right)$ and the variance $\sigma_{\alpha}^2=\int \d\alpha\,\alpha^2 P\left(\alpha\right)$. Since we want to compare results to the box distributed case from section \ref{scenarios}, where the variance is given by $\sigma_{\alpha}^2=\frac{\Delta_{\alpha}^2}{12}$, we define the width of the distribution $\Delta_{\alpha}$ according to this equation. With the help of both benchmarks we are able to compare the distributions with the scenarios of disorder in only one BH parameter, as introduced in section \ref{scenarios}. The mean value of the distribution $P\left(\alpha\right)$ here matches the site independent BH parameters $\alpha=\epsilon,\,J,\,U$ from section \ref{scenarios}, while the width of the distribution $\Delta_{\alpha}=\sqrt{12 \sigma_{\alpha}^2}$ corresponds to the disorder strength given as free parameter in section \ref{scenarios}.\\
The resulting mean value and width of the distribution are shown in Figure \ref{MeanValueWidth}. The amplitude of the main lattice $s_1$ is one order of magnitude larger than the one of the second lattice $s_2$. In a shallow lattice ($s_1$ small) the mean value of the on-site energy $\epsilon$ and the inter-particle interaction $U$ are small and grow with increasing depth of the lattice ($s_1$ large). The mean value of the tunneling rate $J$ reaches its maximal value in a shallow lattice and decreases in a deep lattice. All mean values are independent of the strength $s_2$ of the second lattice. The width of the distribution of the on-site energy $\Delta_{\epsilon}$ is independent of the amplitude $s_1$ of the main lattice, but increases with the amplitude $s_2$ of the second one. As expected, the amplitude $s_2$ of the second lattice indeed increases the disorder strength in the system. The width of the distributions of the tunneling rate $\Delta_{J}$ and the inter-particle interaction $\Delta_{U}$ show similar behavior, depending on both parameters $s_1$ and $s_2$, however, their maximal values differ substantially. Both show increasing widths for increasing $s_2$ and adopt the maximal values for a shallow ($s_1$ small) and strongly disordered ($s_2$ large) lattice.\\
The critical disorder strength at which the last Mott-lobe disappears, is $\Delta_{\epsilon}^c/U=1$ for pure on-site energy disorder and $\Delta_{J}^c/U=\frac{3-2\sqrt{2}}{2}\approx0.0858$ for pure tunneling disorder. Above these vales only the BG and SF phase remain. A comparison of the width of the distribution in units of $U$ (see Figure \ref{MeanValueWidthU}) with the results on disorder in only one BH parameter in section \ref{scenarios} shows that both, the width of the distribution of the on-site energy, as well as the tunneling rate, reaches the region where all three phases occur in the phase diagram. Even though the occurring width of the distribution of the tunneling rate $\Delta_{J}$ is small, it reaches the parameter range, where all three phases compete in the phase diagram. In contrast, the width of the inter-particle interaction $\Delta_U$ is indeed small in comparison to the range in which all three phases occur in the phase diagram and thus may be neglected.  

\section{Phase diagrams of the bichromatic potential}\label{Phase diagrams of the bichromatic potential}
\subsection{Intensity phase diagrams}
\begin{figure}[!htb]
a)\includegraphics[width=3.9cm]{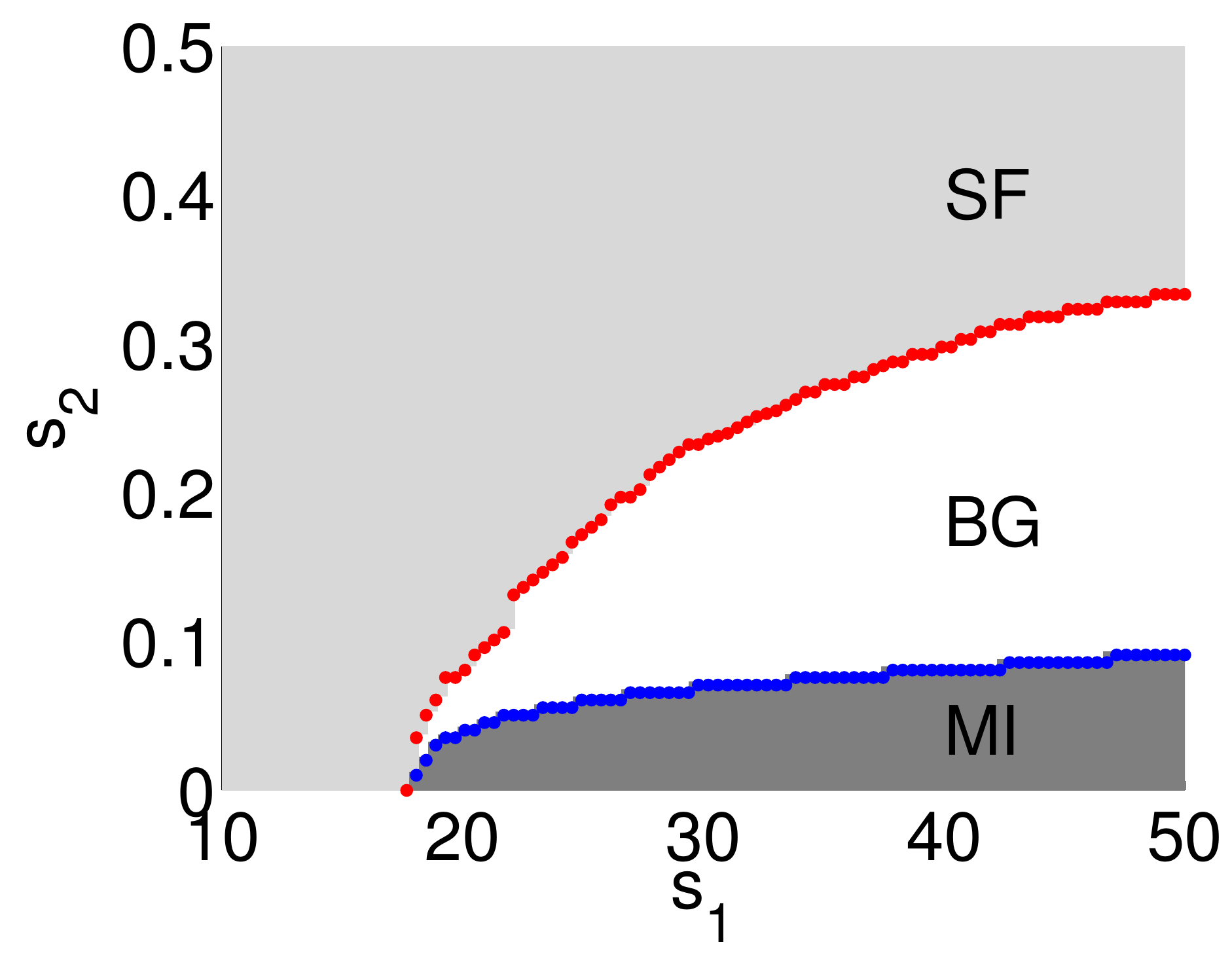}
b)\includegraphics[width=3.9cm]{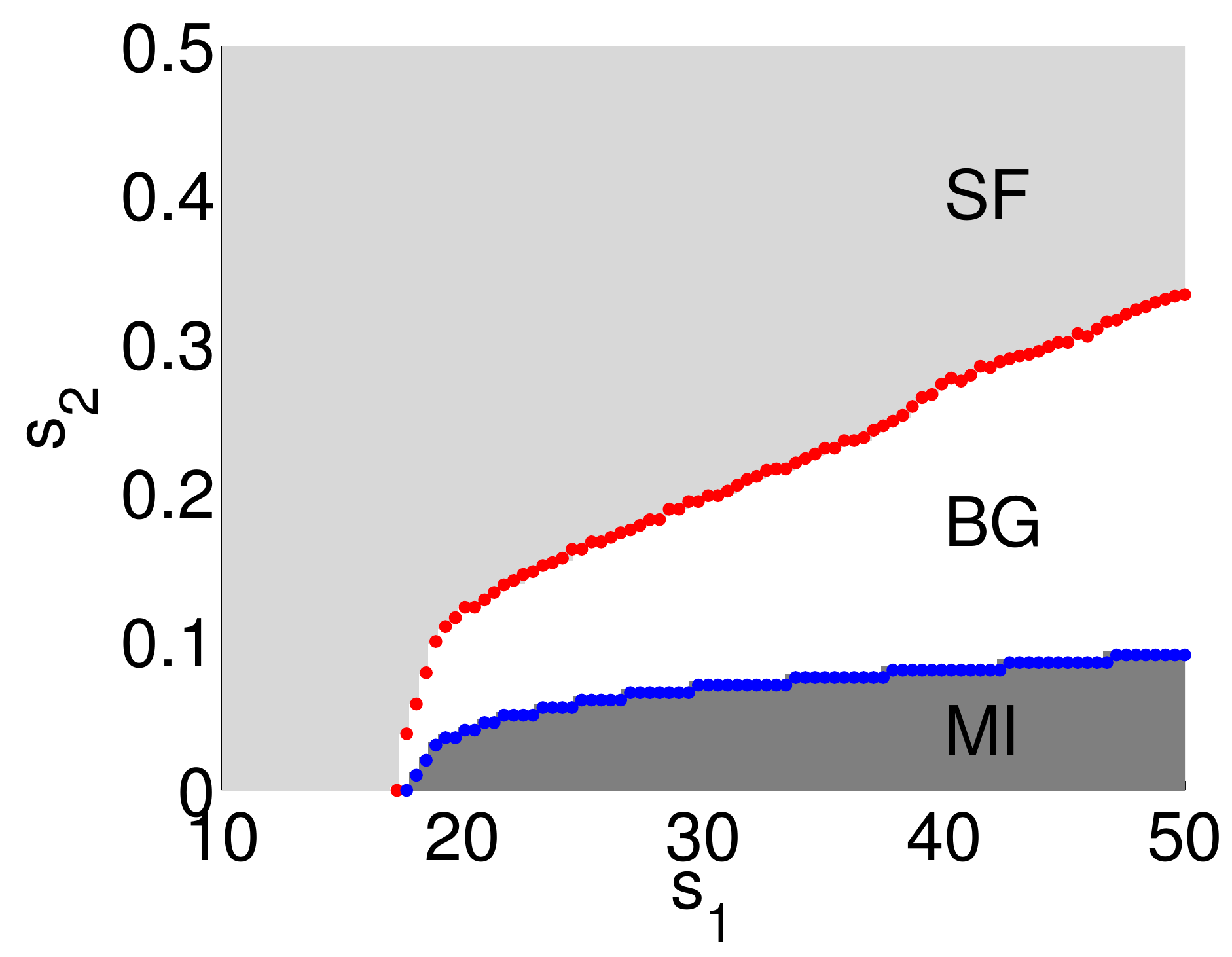}
\caption{\label{Diagram} {\bf Left:} Phase diagram for the bichromatic quasi-periodic potential \eqref{Potential} for particle density $\langle \O[i]n \rangle=1$ in dependence of the laser intensities $s_1$ and $s_2$ for a system with $32\times32$ sites. {\bf Right:} The same phase diagram as on the left side but for a BH model with uncorrelated on-site, hopping strengths and interaction energy disorder with the distribution parameters determined for the bichromatic quasi-periodic potential in section \ref{Distributions}.}
\end{figure}

$\;$ \\
With the help of the determined BH parameters we have determined the phase diagram for fixed average number $N=\sum_i \langle \O[i]n \rangle=M$ of one particle per site in dependence of the laser intensities $s_1$ and $s_1$ by adjusting the chemical potential $\mu$. The resulting phase diagram is shown in Figure \ref{Diagram} a). Here the BH parameters are correlated according to the lattice potential \eqref{Potential}, exemplarily displayed in Figure \ref{Parameter}. According to the LMF cluster analysis described in section \ref{section:model}, the MI phase is characterized by the absence of any SF site, which means that every site in the MI region has an integer particle number. In the SF region sites with non-integer particle number percolate. In between, in the BG region, the system consists of both, sites with integer, and others with non-integer particle number, which do not percolate.\\
Next we compare the obtained phase diagram for the bichromatic quasi-periodic potential with the one obtained for a BH model with uncorrelated disorder according to identical distributions of the BH parameters $P\left(\alpha\right)$ with $\alpha=\epsilon,\,J,\,U$. These distributions depend on both of the laser intensities $s_1$ and $s_2$. We start with one parameter set for $\epsilon$,$J$ and $U$ given by fixed $s_1$ and $s_2$, which we have determined and discussed in sections \ref{Determination} and \ref{Distributions}. We produce $200$ different samples, by randomly choosing new site indices. In other words, we study $200$ samples according to the same distribution by switching lattice sites thereby erasing local correlations in the parameter set. After performing the LMF cluster analysis we determine the BG-SF boundary with finite size scaling. The resulting phase diagram in Figure \ref{Diagram} b) does not differ much from Figure \ref{Diagram} a), only the BG-SF transition line for uncorrelated disorder is slightly distorted in comparison with the quasi-periodic case.\\ 
In the resulting phase diagram, shown in Figure \ref{Diagram} b), all three phases occur in dependence of the lattice parameters $s_1$ and $s_2$. Along the $s_1$-axis ($s_2=0$) the direct SF-MI transition of the ordered system occurs. For values below this point in a shallow lattice the SF phase covers the whole parameter region independent of $s_2$. This corresponds to the fact that in this region the tunneling rate is largest, as shown in Figure \ref{MeanValueWidth}. Above this point the MI occurs, which is completely surrounded by the BG for intermediate $s_2$, which in turn is enclosed by the SF phase for even larger amplitudes of the second lattice $s_2$. Notice that the potential \eqref{Potential} reduces to the ordered case for $s_2=0$ as well as for $s_1=0$. Therefore, along the $s_2$-axis the system also undergoes a direct MI-SF transition. In the region where $s_1\ll s_2$ the second lattice is dominant and a similar structure occurs.

\subsection{BH parameter phase diagrams}
\begin{figure}[!htb]
a)\includegraphics[width=3.9cm]{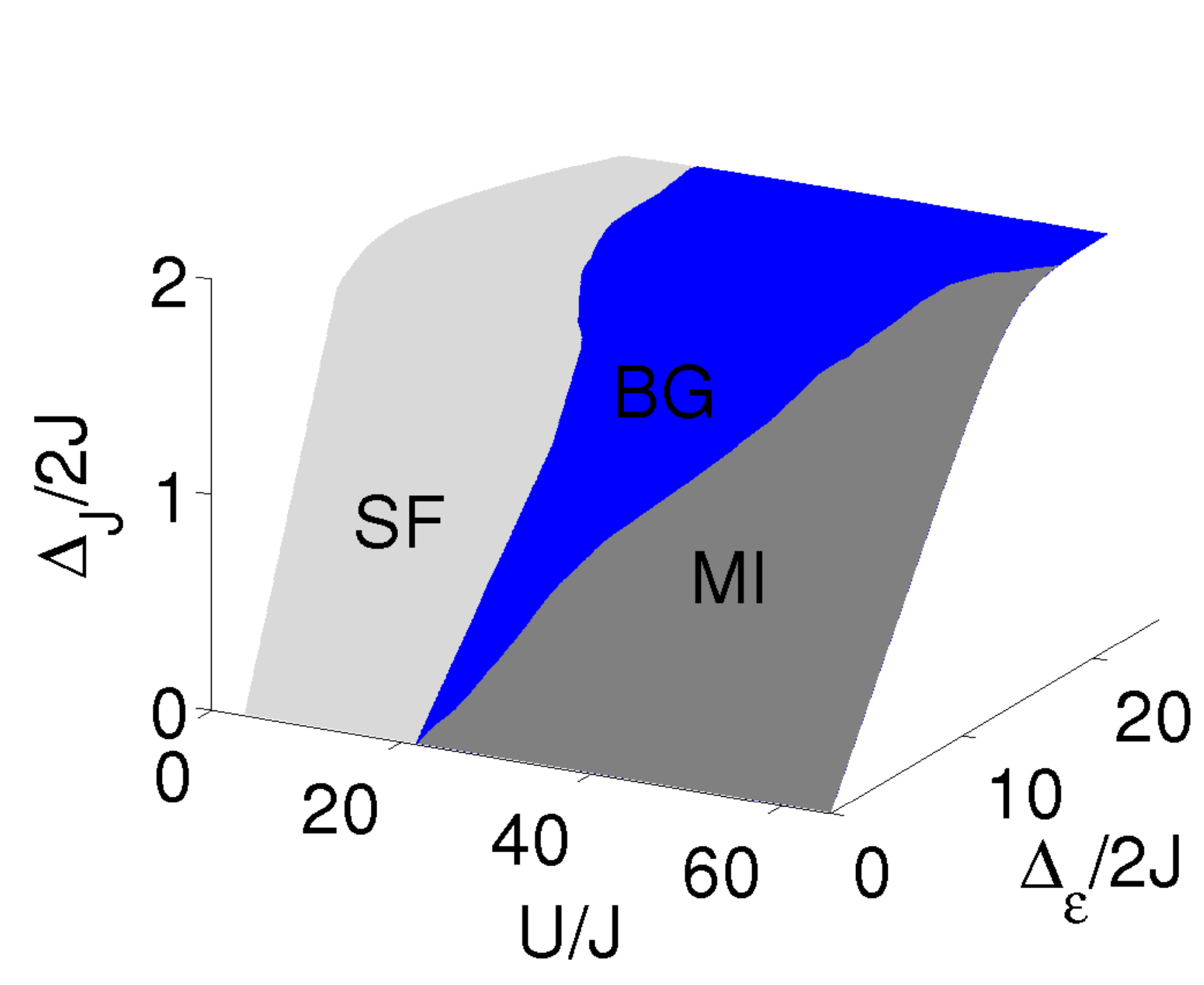}
b)\includegraphics[width=3.9cm]{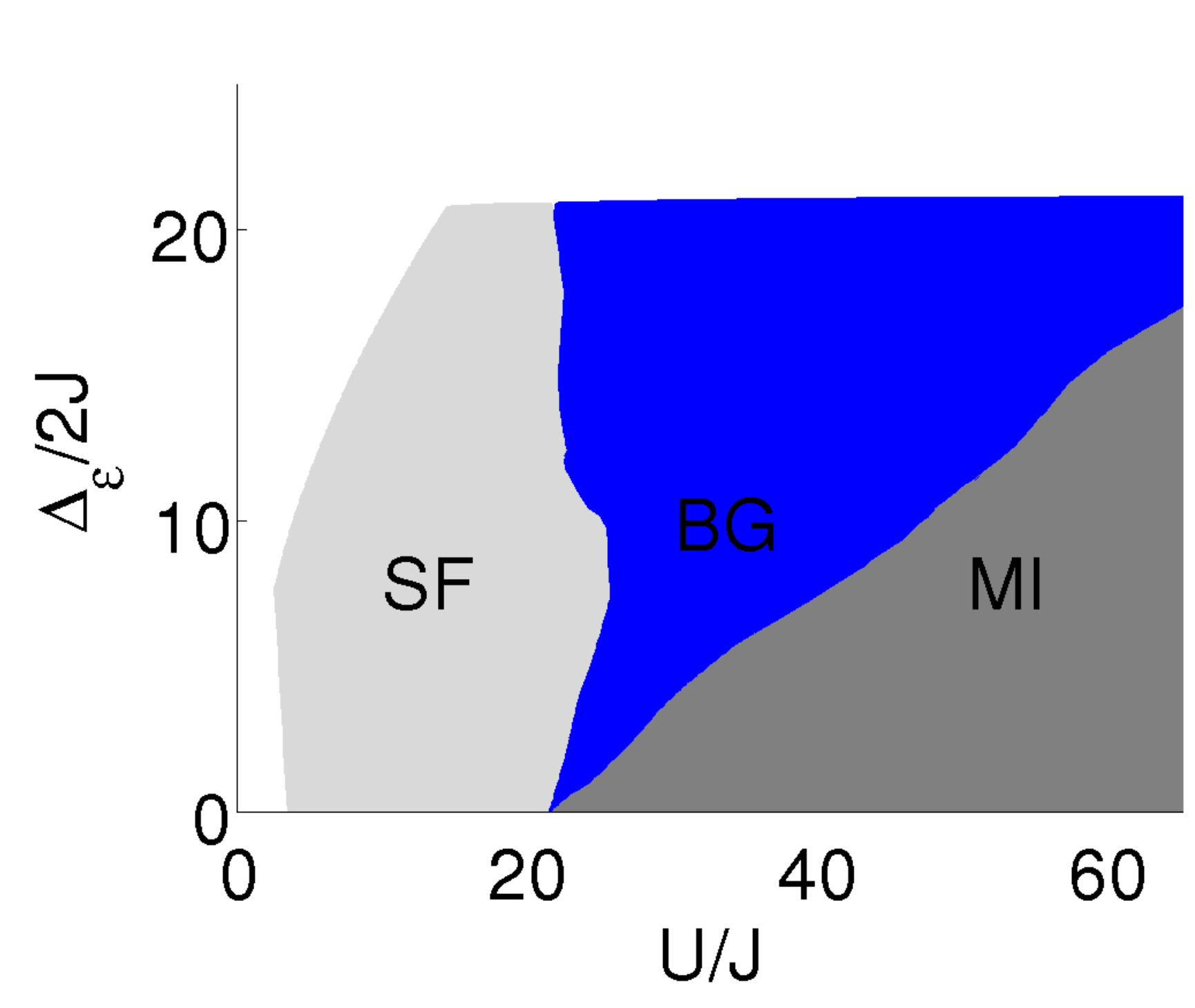}
\caption{\label{DiagramALL} Phase diagram for a quasi-periodic potential \eqref{Potential} in dependence of the BH Parameter.}
\end{figure}

\begin{figure}[!htb]
\begin{tabular}{ccc}
a)\includegraphics[width=2.45cm]{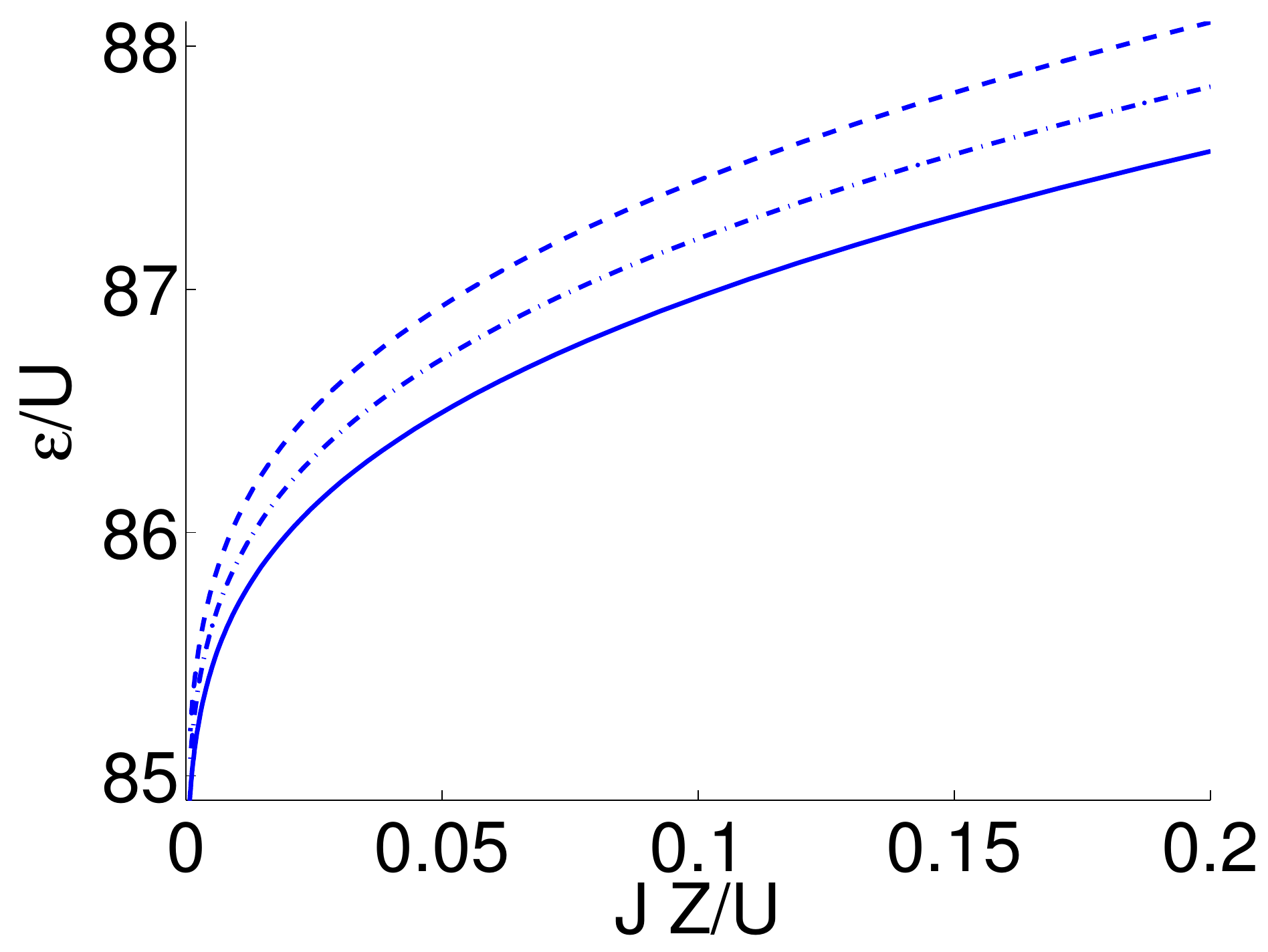} & b)\includegraphics[width=2.45cm]{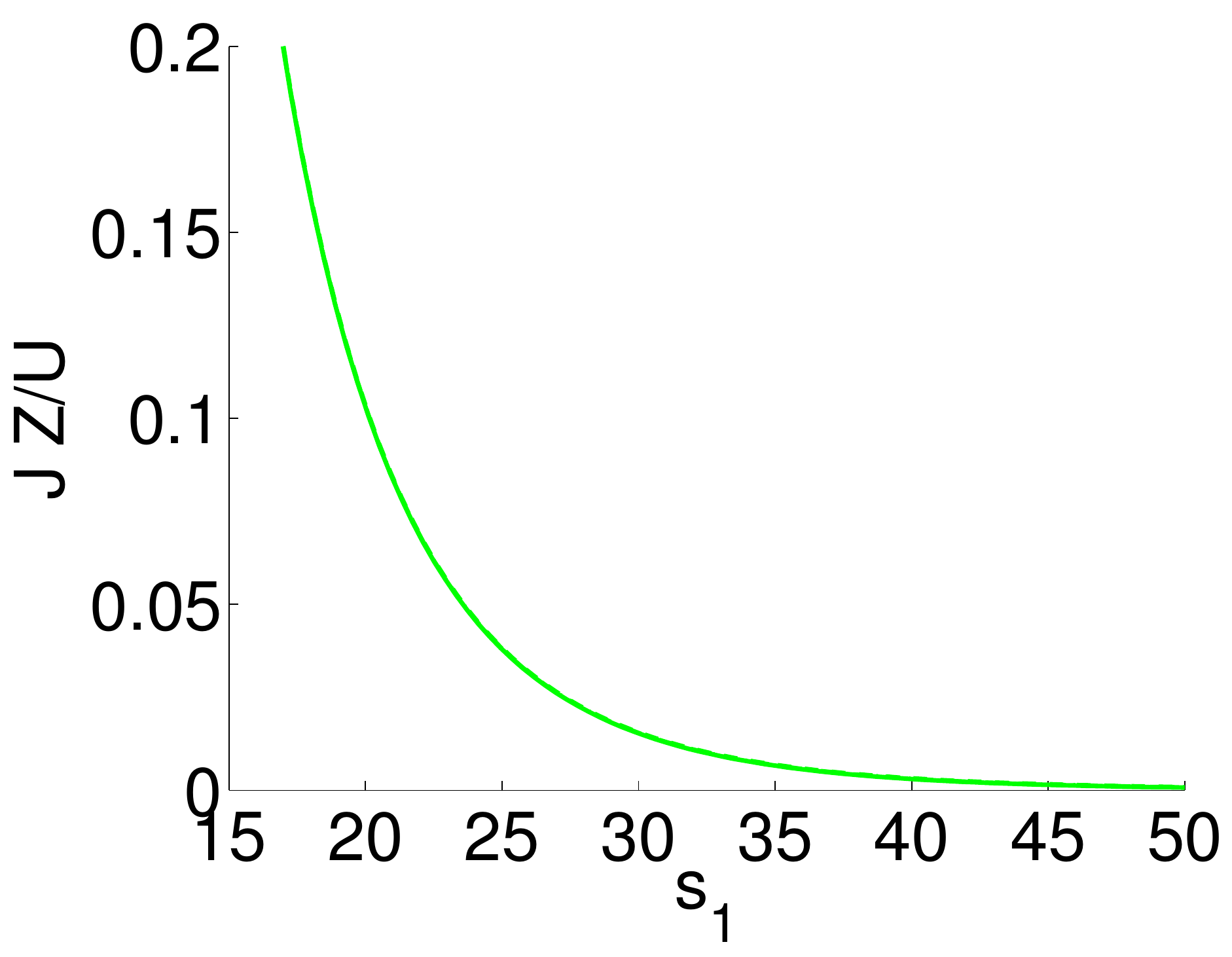} &\\
c)\includegraphics[width=2.45cm]{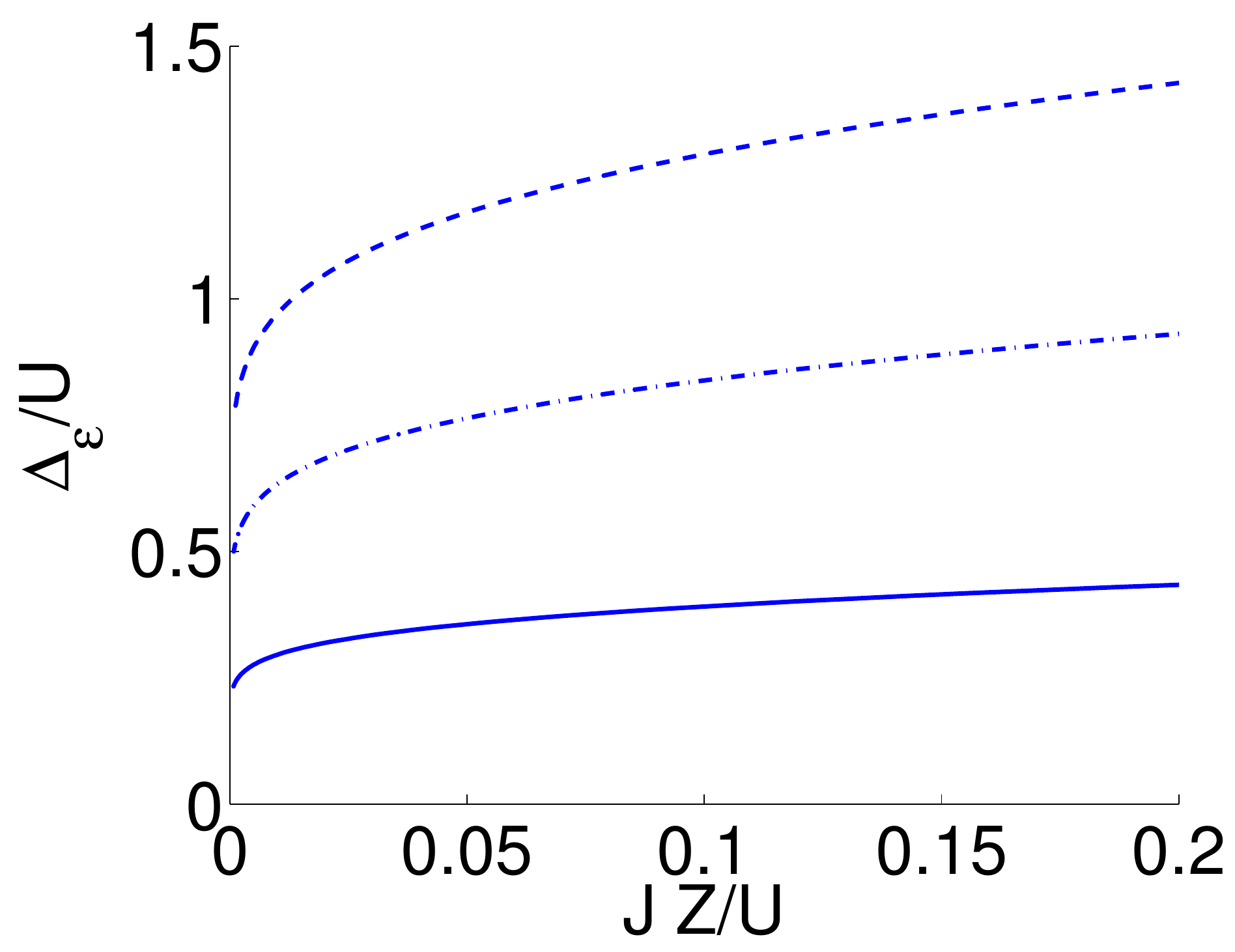} & d)\includegraphics[width=2.45cm]{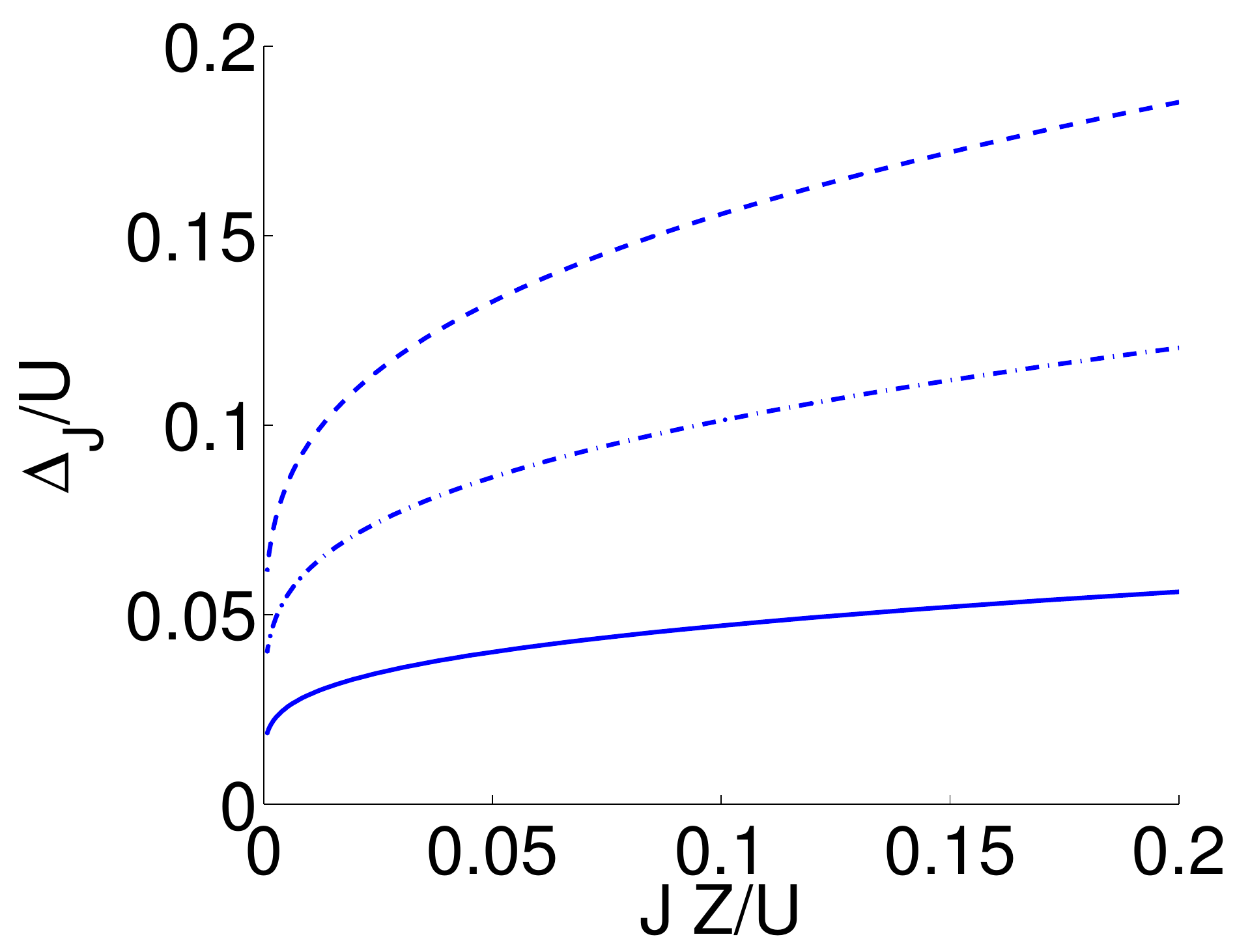} & e)\includegraphics[width=2.45cm]{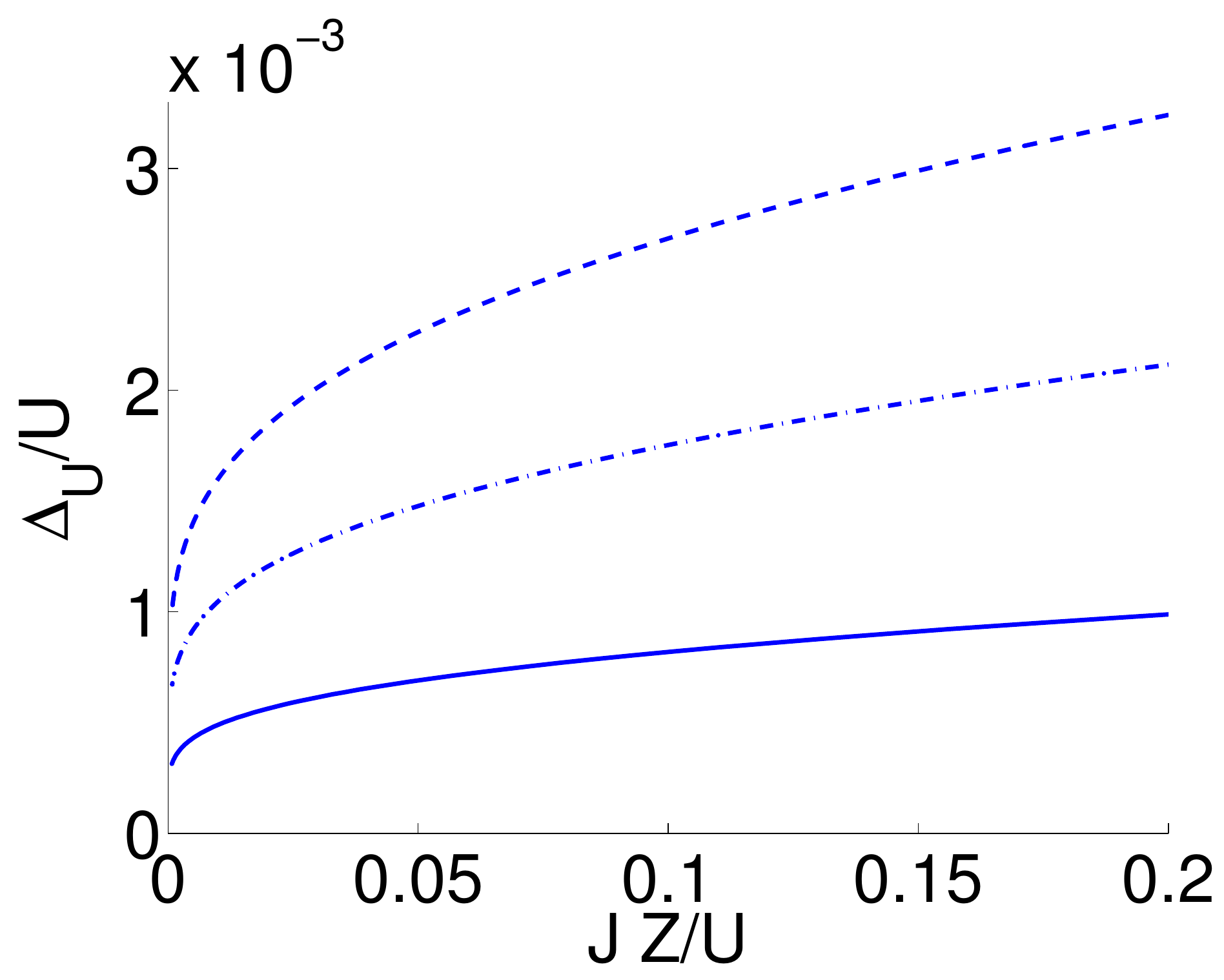}
\end{tabular}
\caption{\label{MeanValueWidthTest} The on-site energy $\epsilon/U$ and the widths $\Delta_{\epsilon}/U,\,\Delta_{J}/U,\,\Delta_{U}/U$ of the BH parameter for different values of $s_2=0.0354\left(\textcolor{blue}{-}\right),\,0.0758\left(\textcolor{blue}{-\cdot}\right),\,0.1162\left(\textcolor{blue}{--}\right)$ as a function of $JZ/U$. In the middle on the top the tunneling rate $JZ/U$ is shown as a function of $s_1$. The tunneling rate is independent of $s_2$ and a unique function of $s_1$. The on-site energy $\epsilon/U$ as well as all widths increase with the tunneling rate $J Z/U$ and  the amplitude $s_2$.}
\end{figure}

\begin{figure}[!htb]
a)\includegraphics[width=3.9cm]{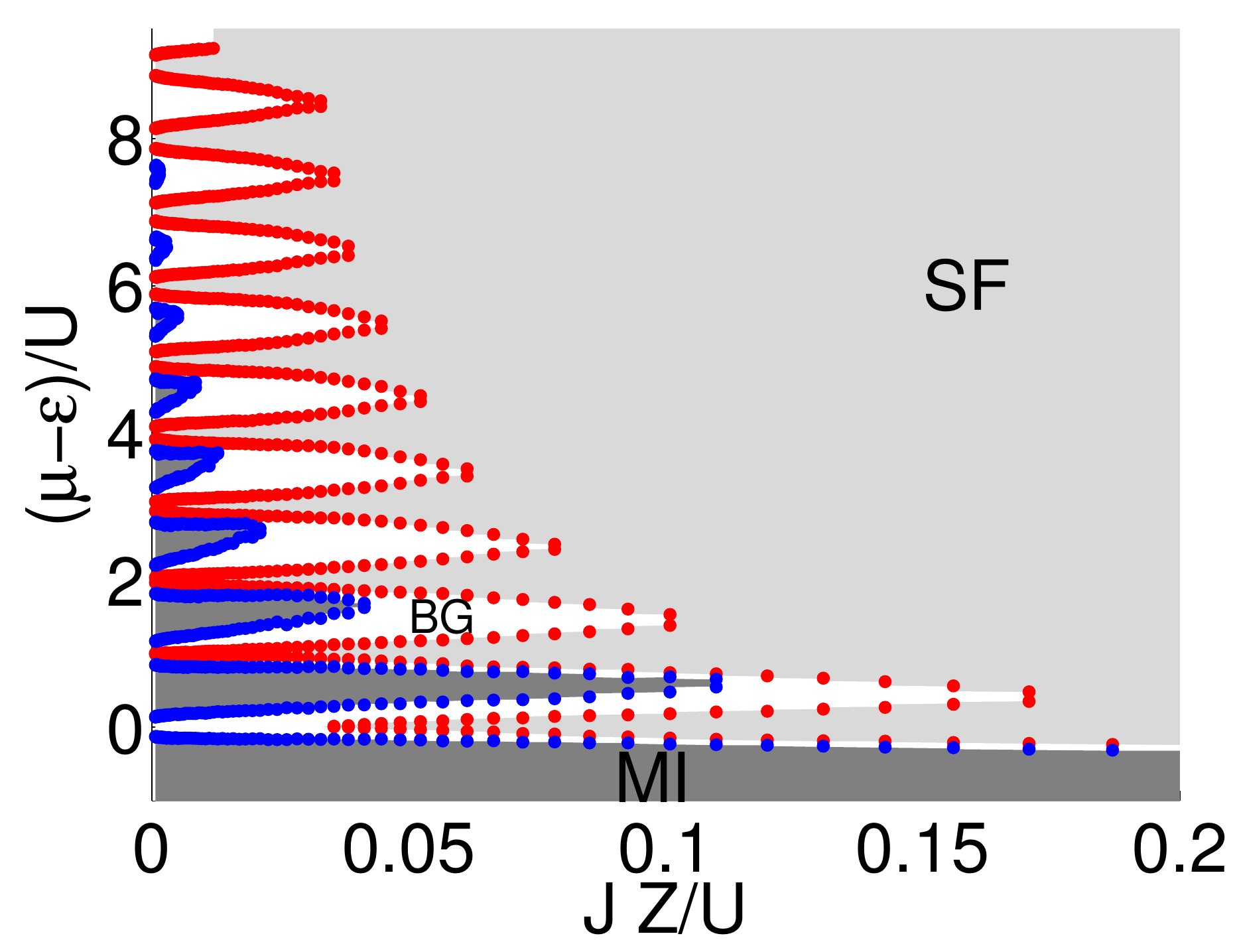}
b)\includegraphics[width=3.9cm]{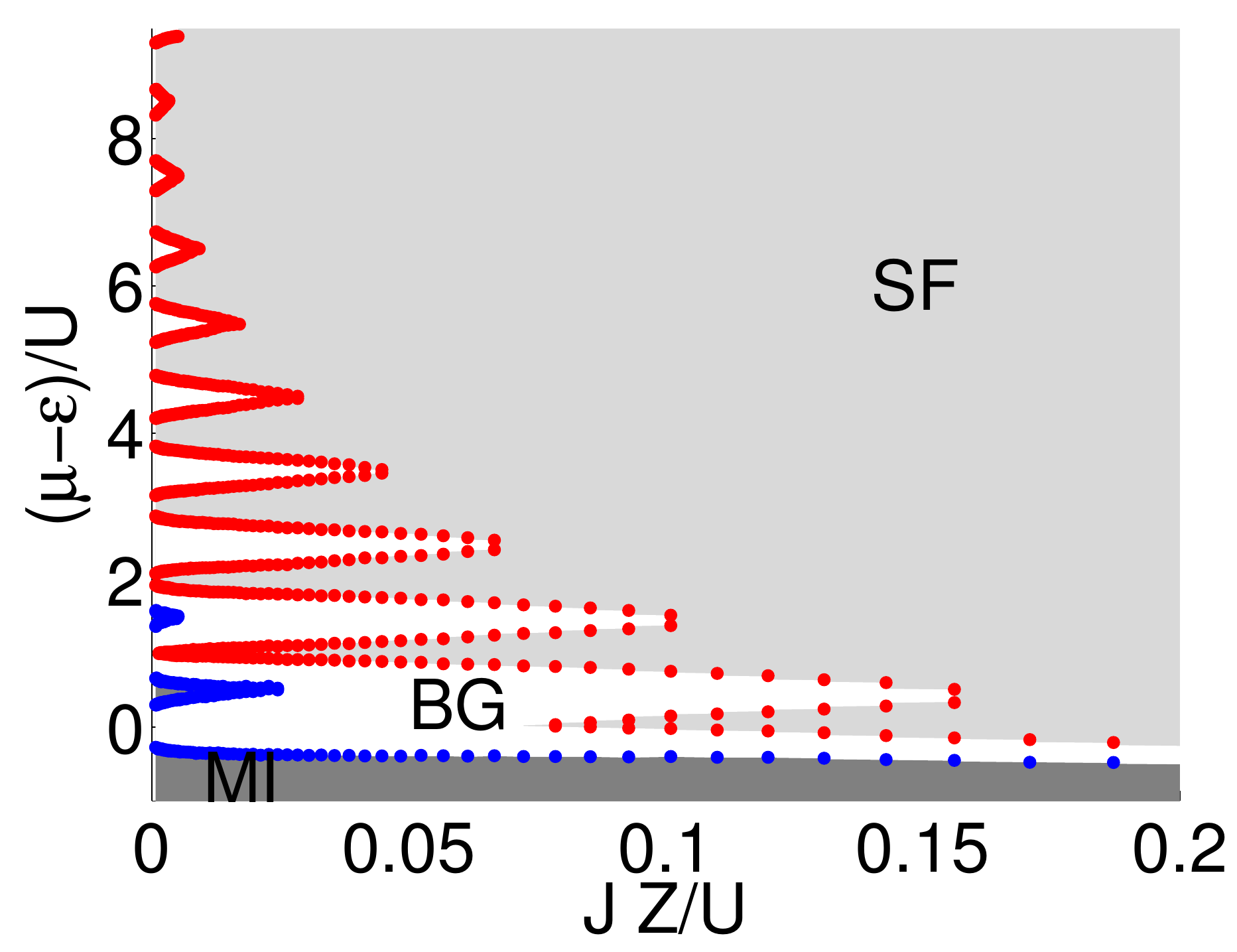}
c)\includegraphics[width=3.9cm]{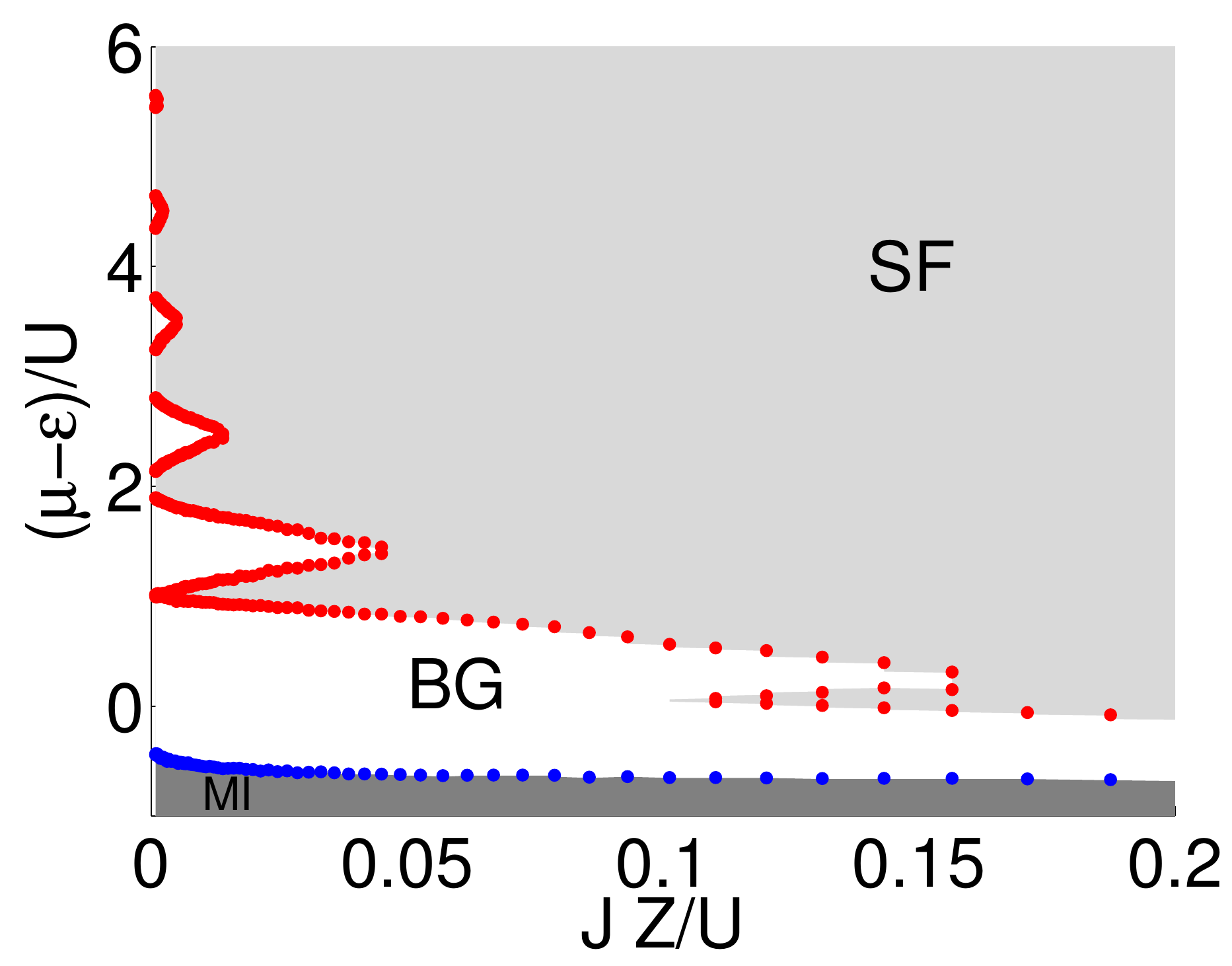}
\caption{\label{PhasenTest} Phase diagram for different \mbox{$s_2=0.0354,\,0.0758,\,0.1162$} as a function of the mean tunneling rate $J Z/U$ and the chemical potential $\left(\mu-\epsilon\right)/U$.}
\end{figure}

$\;$ \\
Let us now have a look at the phase diagram in dependence of the BH parameters. In the case of only one disordered BH parameter two possible representations of the phase diagram are common: For the first one \cite{Soey11,Nied13} the particle number is fixed to one particle per site, which fixes the chemical potential $\mu$. Then, the phase transitions are shown in dependence of the inter-particle interaction $U/J$ and the disorder strength $\Delta/2J$. The second representation \cite{Buon09,Buon07,Buon04b,Nied13} shows the Mott-lobes in dependence of the tunneling rate $J Z/U$ and the chemical potential $\mu/U$ for fixed disorder strength $\Delta/U$. In this section we will discuss our results in both representations, keeping in mind that all BH parameters as well as all disorder strengths are functions of $s_1$ and $s_2$, and thus are not independent of each other.\\
The data from the $\left(s_1,s_2\right)$-phase diagram shown in Figure \ref{Diagram} can be translated into a diagram similar to the first representation: According to Figure \ref{MeanValueWidth} the mean parameters $\epsilon$, $J$, $U$ as well as the widths $\Delta_{\epsilon}$, $\Delta_{J}$, $\Delta_{U}$ are functions of the two amplitudes $s_1$ and $s_2$. Since $\Delta_{U}$ is two orders of magnitude smaller than the other widths, the inter-particle interaction $U$ can be treated as a sharp value to a good approximation. As a result the phase diagram can be visualized as the surface $\left(U/J,\,\Delta_{\epsilon}/2J,\,\Delta_{J}/2J \right)$ in three dimensions. This is shown in Figure \ref{DiagramALL}, where each phase is colored differently. Notice that with a quasi-periodic potential \eqref{Potential}, which depends on the two amplitudes $s_1$ and $s_2$, only this surface in the BH parameter space can be reached, since all BH parameters are functions of $s_1$ and $s_2$ and dependent on each other. As a consequence disorder, e.g, where only one parameter is disordered while the others are fixed, cannot be reached in the phase diagram. Either it is an ordered $\left( \Delta_{\epsilon}=\Delta_{J}=0 \right)$ or a completely disordered $\left( \Delta_{\epsilon}\neq0,\,\Delta_{J}\neq0\right)$ system. This has two important implications: With a quasi-periodic potential neither the whole parameter space nor a pure on-site disorder can be realized.\\      
The phase diagram in Figure \ref{DiagramALL} shows all three phases: The BG phase (blue) separates the MI (dark gray) phase at strong inter-particle interaction $U/J$ from the SF regime (light gray) at weak inter-particle interaction $U/J$. In Figure \ref{DiagramALL} b) the same phase diagram is show as a projection on the $\left(U/J,\,\Delta_{\epsilon}/2J\right)$-plane.\\
The phase boundaries differ substantially from those of the BH model with uncorrelated disorder exclusively in the on-site energies \cite{Soey11,Nied13} due to the additional presence of disorder in the hopping strengths. More drastic is the difference between Figure \ref{DiagramALL} and the $V_2/J$-$U/J$ phase diagram predicted for a one-dimensional BH-model with bichromatic quasi-periodicity only in the on-site potential \cite{Roux08}. Since according to Figure \ref{MeanValueWidth} d) $\Delta_\epsilon$  is proportional to $V_2=s_2 E_{R2}$, the phase diagram in Figure \ref{DiagramALL} is directly comparable to Figure 1 (left) of \cite{Roux08} and Figure 3 (bottom) of \cite{Deng08}, both of which show that a direct MI-SF transition occurs. The latter is absent in Figure \ref{DiagramALL}, where an intervening BG phase occurs between the MI and SF phase. This might be explained by the fact that already a small amount of disorder in the hopping strengths strongly enlarges the BG regions in the phase diagram, as Figure \ref{PhasenJ} demonstrates.\\
For the second representation we fix the weaker amplitude $s_2$, which introduces disorder to the system, and study the system in dependence of $s_1$ and $\mu$. Since the tunneling rate $J$ is a unique function of $s_1$ and independent of $s_2$, as shown in Figure \ref{MeanValueWidthTest} b), the $s_1$-axis can easily be converted into a $J$-axis. In theoretical works disorder is usually introduced by bounded distributions with zero mean values. In the quasi-random case the mean value of the distributions $P\left(\alpha\right)$ of the BH- parameters $\alpha=\epsilon,\,J,\,U$ are non-zero, as shown in Figure \ref{MeanValueWidth}. In order to take this into account, we use $\mu-\epsilon$ instead of simply $\mu$. Thus, from the data in the $\left(s_1,\,\mu\right)$-plane, we can extract a phase diagram in dependence of $J Z/U$ and $\left(\mu-\epsilon\right)/U$, as shown in Figure \ref{PhasenTest}. Notice that $J,\,U,\,\epsilon$ as well as $\Delta_{J},\,\Delta_{U},\,\Delta_{\epsilon}$ are all functions of $s_1$ and $s_2$ and not independent from each other, as they are asumed to be in most simulations of disordered systems.\\
In Figure \ref{MeanValueWidthTest} the behavior of the system parameters for different values of $s_2=0.0354,\,0.0758,\,0.1162$ is shown as a function of $J Z/U$. Figure b) shows the tunneling rate in dependence of $s_1$. While the tunneling rate $J Z/U$ is independent of $s_2$, the on-site energy $\epsilon/U$ varies for different values of $s_2$. In a shallow lattice ($s_1$ small) the tunneling rate $J Z/U$ is large, while for increasing $s_1$ it approaches zero and finally in a deep lattice ($s_1$ large) the tunneling rate $J Z/U$ becomes infinitesimally small. This means that in the phase diagram the $\mu/U$-axis at $J Z/U=0$ may be approached with arbitrary accuracy, but can never be reached. In a deep lattice ($J Z/U$ small, $s_1$ large) the on-site energy $\epsilon/U$ is small and increases with growing $J Z/U$. The disorder strengths $\Delta_{\epsilon}/U,\,\Delta_{J}/U,\,\Delta_{U}/U$ increase with $J Z/U$ and the amplitude $s_2$.\\ 
The phase diagram for different values of $s_2$ as a function of $J Z/U$ and $\left(\mu-\epsilon\right)/U$ is shown in Figure \ref{PhasenTest}. Notice that the BH parameters and their widths vary along the $J Z/U$-axis corresponding to Figure \ref{MeanValueWidthTest}. For all values of $s_2$ we find a regular structure of Mott-lobes, surrounded by individual BG regions. The number of Mott-lobes as well as the number of BG regions decreases with increasing disorder amplitude $s_2$, which is equal to the increase of the disorder strengths $\Delta_{\epsilon}/U,\,\Delta_{J}/U,\,\Delta_{U}/U$. The regular pattern of Mott-lobes and BG regions repeats itself in intervals of length one along the $\left(\mu-\epsilon\right)/U$-axis. The lower and the upper extent of the Mott-lobes have the same distance to the next integer number for fixed $s_2$. Thus, the Mott-lobes have the same width along the $\left(\mu-\epsilon\right)/U$-axis, while their extension in $J Z/U$-direction shrinks with their number $n$. Except of the first BG region all the others are separated from each other by SF regions, reaching down to very small tunneling rates $J Z/U$. This is a unique feature of disorder only in the tunneling rates, which was discussed in section \ref{section:DissJ}. The fact that we see this special phenomenon here in the phase diagram of a quasi-random potential, once more promotes our finding from section \ref{Distributions} that the influence of disorder in the tunneling rate cannot be neglected.

\section{Conclusion}\label{section:Conclusion}
The main result of this paper is that disorder in the tunneling rate must be taken into account in setups using 2d bichromatic quasi-periodic potentials. It influences the phase diagram just as much as the on-site disorder. This claim is supported by several findings in this paper:\\
Firstly, in section \ref{scenarios} we showed that the disorder strength, where all three phases compete in the phase diagram, is one order of  magnitude smaller for tunneling disorder than for on-site interaction. We found that each scenario, in which only one BH parameter is disordered, yields different features in the phase diagram. Especially, the characteristics of tunneling disorder are a finite number of Mott-lobes and the existence of SF regions even for $J Z/U=0$.\\
Secondly, we discussed bichromatic quasi-periodic potentials in section \ref{quasi-periodic} and showed (see Figure \ref{MeanValueWidth}) that the width of the distribution of the tunneling rate as well as that of the on-site energy reach the physical interesting region, where all three phases compete. This is true even though the width of the distribution of the tunneling rate is one order of magnitude smaller than that of the on-site energy. The influence of disordered inter-particle interaction is negligible, since its width remains four orders of magnitude below the critical disorder strength. This is in agreement with the results for distributions of BH parameters produced by a random diffuser pattern overlapping the main lattice \cite{Whit09,Zhou10}. Correspondingly to our work, in these papers the authors also showed, that the width of the tunneling rate and the inter-particle interaction are several orders of magnitude smaller than that of the on-site energy. Moreover, it is important to keep in mind that in a bichromatic quasi-periodic potential it is not possible to study exclusively on-site disorder. For growing intensity of the second laser $s_2$, the widths of the disorder distributions of both, the on-site, and the tunneling rate, increase simultaneously.\\
Thirdly, the influence of tunneling disorder is obvious in the phase diagram of the quasi-periodic potential in dependence of the BH parameters. The transition lines, shown in Figure \ref{DiagramALL}, deviate from pure box distributed on-site disorder, discussed for example in \cite{Nied13,Soey11}. In the quasi-random case the SF region is smaller, while the BG and the MI regions cover a larger region. In the $\mu$-$J$-phase diagram of Figure \ref{PhasenTest} we find individual BG regions, which are separated by SF regions, which is a unique feature exclusively occurring in systems with disordered tunneling rates (see Figure \ref{PhasenJ}).\\
While the field of box on-site disorder was studied widely \cite{Soey11,Lin11,Capo08,Poll09,Prok04,Lee01,Kisk97,Kisk97a,Krau91,Roux08, Deng08, Deng09,Carr10,Raps99,Free96,Shes93,Buon09,Buon07,Buon04b,Nied13,Biss09, Biss10}, the works on tunneling disorder are rare and mainly deal with bimodal distributions \cite{Prok04,Bala05,Seng07}. A phase diagram far above the critical disorder strength $\Delta_{J}^c/U\approx8.58\,10^{-2}$ for equally distributed tunneling disorder is shown in \cite{Biss10}.\\
The one-dimensional BH model with a bichromatic on-site potential with incommensurable wave lengths was studied in \cite{Roux08,Deng08}. The motivation there was, as also in this paper, to qualitatively understand the phase diagram of bosons in a bichromatic quasi-periodic potential, but it was argued that the variations in the hopping strengths as well as in the in the interaction energies were only minor and could be neglected. Although, we have shown here by explicit calculation that the disorder strength in the tunneling strengths is indeed one order of magnitude smaller than the on-site disorder, it nevertheless has a strong effect on the 2d phase diagram. The most striking difference is that the phase diagrams of \cite{Roux08,Deng08} show a direct MI-SF transition (which for quasi-period disorder is not in contradiction with general predictions for uncorrelated disorder \cite{Poll09}), whereas we find an intervening BG phase between the MI and the SF phase for all values of the laser intensities $s_1$ and $s_2$. A reason for this, in addition to potentially qualitative differences between one and two dimensions, could be that our results for uncorrelated disorder in section \ref{scenarios} show that a modest amount of disorder in the hopping strength already generates relatively large BG regions in the phase diagram (see \mbox{Figure \ref{PhasenJ}}). Consequently, disorder in the hopping strengths cannot be neglected studying bosons in a bichromatic quasi-periodic potential.\\
Conversely, one should be aware that experimental realizations of the disordered potential by a bichromatic quasi-period potential, as in \cite{Lye05,Fall07}, produce a phase diagram that is qualitatively very different from the predictions of the disordered BH model with exclusively on-site disorder.
\begin{acknowledgments}
We thank G. Morigi for fruitful and stimulating discussions and acknowledge financial support from the German Research Foundation DFG under grant number GRK 1276.
\end{acknowledgments}

\end{document}